\documentclass[12pt]{article}
\usepackage[a4paper,left=3cm,right=3cm,top=3cm,bottom=3cm]{geometry}
\usepackage[utf8]{inputenc}
\usepackage{lmodern}
\usepackage[onehalfspacing]{setspace} 
\usepackage[T1]{fontenc} 
\usepackage[UKenglish]{babel} 
\usepackage[UKenglish]{isodate}
\usepackage{float} 
\usepackage{cancel} 
\usepackage{hyperref}
\hypersetup{
  colorlinks   = true, 
  urlcolor     = NavyBlue,
  linkcolor    = NavyBlue, 
  citecolor   = NavyBlue 
}
\usepackage[dvipsnames]{xcolor}
\usepackage{scalerel}
\usepackage[authoryear, comma]{natbib}
\usepackage{chapterbib}
\usepackage{amsmath} 
\usepackage{amssymb} 
\usepackage{amsfonts} 
\usepackage{amsthm} 
\usepackage{bbm}
\usepackage{comment}
\usepackage{txfonts}
\usepackage{chngcntr}
\usepackage{apptools}
\usepackage[printonlyused]{acronym} 
\usepackage{url}

\usepackage{setspace}

\def\a{\alpha}
\def\b{\beta}

\def\e{\varepsilon}

\def\l{\lambda}
\def\m{\mu}

\def\w{\omega}

\def\W{\Omega}

\def\N{\mathbf{N}}

\def\BB{\mathcal{B}}

\def\NN{\mathcal{N}}

\DeclareMathOperator{\E}{\mathbb{E}}

\DeclareMathOperator{\Var}{Var} 

\DeclareMathOperator{\sgn}{sgn} 

\newcommand{\Abs}[1]{\left\lvert #1 \right\rvert}

\newcommand{\Paren}[1]{\left( #1 \right)}

\newcommand{\Brac}[1]{\left[ #1 \right]}

\newcommand{\de}{\mathop{}\!\mathrm{d}}

\newtheorem{Thm}{Theorem}
\newtheorem{Ass}{Assumption}
\newtheorem{Pro}{Proposition}

\newtheorem{Cor}{Corollary}
\newtheorem{Rem}{Remark}

\newtheorem{Def}{Definition}

\newtheorem{Claim}{Claim}
\theoremstyle{definition}
\usepackage{sgame} 
\usepackage{subfig} 
\usepackage{tikz}
\usepackage{tkz-euclide}
\usepackage{graphicx}
\usetikzlibrary{calc,trees,positioning,arrows,chains,shapes.geometric,%
    		decorations.pathreplacing,decorations.pathmorphing,shapes,%
    		matrix,shapes.symbols}
    	\tikzset{
    		>=stealth',
    		punktchain/.style={
    			rectangle, 
    			rounded corners, 
    			draw=black, very thick,
    			text width=10em, 
    			minimum height=3em, 
    			text centered, 
    			on chain},
    		line/.style={draw, thick, <-},
    		element/.style={
    			tape,
    			top color=white,
    			bottom color=blue!50!black!60!,
    			minimum width=8em,
    			draw=blue!40!black!90, very thick,
    			text width=10em, 
    			minimum height=3.5em, 
    			text centered, 
    			on chain},
    		every join/.style={->, thick,shorten >=1pt},
    		decoration={brace},
    		tuborg/.style={decorate},
    		tubnode/.style={midway, right=2pt},
    	}

\makeatletter    
\newcommand*{\defeq}{\mathrel{\rlap{%
			\raisebox{0.3ex}{$\m@th\cdot$}}%
		\raisebox{-0.3ex}{$\m@th\cdot$}}%
	=}
\makeatother

\newtheoremstyle{break}
  {\topsep}{\topsep}%
  {\itshape}{}%
  {\bfseries}{}%
  {\newline}{}%
\theoremstyle{break}

\begin{document}
    	\begin{titlepage}

\cleanlookdateon

		\title{Dynamic Competition for Attention}
		\author{Jan Knoepfle\thanks{School of Economics and Finance, Queen Mary University of London,  \href{mailto:J.knoepfle@qmul.ac.uk}{j.knoepfle@qmul.ac.uk}. For invaluable feedback and support, I am indebted to Marina Halac, Johannes  H\"orner, Stephan Lauermann 
and Sven Rady.  I thank Zeinab Aboutalebi, David Ahn,  Ian Ball, Dirk Bergemann, Francesc Dilm\'e, Thomas Graeber, Daniel Hauser, Ryota Iijima, Deniz Kattwinkel, Patrick Lahr, Igor Letina,  Niccol\`o Lomys, Pauli Murto, Aniko \"Ory, Anna Sanktjohanser, Eran Shmaya, Ludvig Sinander, Andre Speit, Philipp Strack, Yiman Sun, Juuso V\"alim\"aki, Allen Vong and Weijie Zhong; as well as numerous audiences 
for insightful conversations and suggestions. 
I am grateful to the Yale Economics Department for their hospitality during my visit where this project was initiated. 
This work was supported by the German Research Foundation (DFG) through CRC TR 224 (Project B04) and by the Academy of Finland (project number 325218).
		}}
		
		\date{\today}

		\maketitle \thispagestyle{empty} 

	\begin{center}
	    \large{ }
	\end{center} 
			\vspace*{1cm}
		\begin{abstract}
			\noindent 
		This paper studies information transmission from multiple senders who compete for the attention of a decision maker. Each sender is partially informed about the state of the world and decides how to reveal her information over time to maximise attention. A decision maker wants to learn about the state but faces an attention cost.
We derive a condition on the informational environment and the decision problem that guarantees that all information from the senders can be transmitted to the decision maker in equilibrium. 
A simple class of information processes implements full transmission across general environments. The attention each sender receives is proportional to the residual value of her information.
In the case of conditionally iid-informed senders, in the limit as the number of senders grows large, the receiver learns the state exactly and immediately (at no attention cost).
		\end{abstract}
	
	\end{titlepage}
	\setcounter{page}{2}
\section{Introduction}

The exchange of information for attention is an increasingly important transaction in modern economies. 
Online content -- such as news, professional product reviews, or weather forecasts -- is offered free of charge, and platforms exploit the attracted attention to create revenue, primarily through advertisements.
In the offline world, consultants offer advice with the aim of being consulted as often as possible, and within the firm, department heads want to be consulted and have the ear of the CEO. 
 This paper studies the transmission of information to a receiver when multiple senders compete for the receiver's attention. 

The exchange of information for attention is a complex transaction. Firstly, when maximising attention -- in contrast to financial returns -- senders cannot ask for a large (monetary) price in exchange for the immediate revelation of a large amount of information. 
Rather, they need to collect attention \textit{gradually} by engaging the receiver repeatedly, potentially facing, at each time, the competition of other senders offering their information.  
Secondly, the \textit{design choice} of senders is extremely rich: rather than choosing a one-dimensional parameter determining quantity or quality, the space of revelation policies of each sender's information is infinite-dimensional and could involve complex intertemporal links.\footnote{For example, a sender may first reveal her information in some encrypted way, and then disclose the decryption key at a later time.}
Finally, there are important externalities: the information observed from one sender changes the receiver's valuation for further information as well as his probability assessment of the information other senders can offer.

This paper presents a framework that allows for a tractable analysis of this information-attention exchange.
Addressing the externalities, we derive a \textit{substitutes condition} on the externalities between senders' informational endowments under which all information can be transmitted to the receiver at maximal speed.\footnote{This condition is related to the Gross Substitute (GS) conditions in \cite{kelso1982job} and \cite{gul1999walrasian}, adapted to our setting where the traded good is information.} 
Despite the rich space of policies for the senders, we identify a simple class of information offers that implement full information transmission across all specifications that satisfy the substitutes condition.

The model features multiple senders and one receiver. The receiver has to take an action and wants to learn about an unknown state to maximise his utility. Senders maximise the attention attracted from the receiver.
	Each sender is endowed with partial information about the state, which we call this sender's component. Information is transmitted in repeated rounds, each round representing one unit of the receiver's attention. 
	At the beginning of each round, each sender offers an arbitrary statistical experiment about her component.\footnote{That is, a distribution over messages, conditional on the realisation of her component.} Senders cannot commit across rounds. The receiver observes all offers and then he either pays an attention cost to consult the experiment of one chosen sender and continue to the next round, or he stops learning and takes the optimal action with his current information. 

The first result, Theorem \ref{thm:monopoly}, characterises the unique equilibrium payoff set when there is only one sender. 
The monopolistic sender always reveals all her information and extracts all surplus from the receiver:  
the receiver's total expected attention cost is equal to the difference in expected utility from taking the action with all information or no information. 
This implies that the sender incurs no loss from her lack of intertemporal commitment; the gradual nature of attention does not reduce the monopolist’s payoff.
However, the lack of intertemporal commitment \textit{does} restrict the set of feasible information-transmission processes: the sender must offer positive value at all times to commit herself, in expectation, to faster information delivery in the future. 
A simple class of sender strategies that achieves this consists of \textit{
All-or-Nothing} (AoN) offers. In this class, the sender posts a probability with which the experiment reveals her available information fully; and with the remaining probability, the experiment delivers no information.
AoN offers reveal information in jumps. We show in a case of finitely many receiver actions, that the lack of intertemporal commitment makes jumps in the revelation process necessary, even in the continuous-time limit.





With multiple senders, due to externalities, and without prices or long-term contracts, the existence of an efficient equilibrium is not a foregone conclusion.\footnote{See also \cite{kelso1982job} and \cite{gul1999walrasian} where externalities prevent existence in other settings. We show by example how externalities can impede information transmission entirely.}
We derive a substitutes condition on the senders' components requiring that any sender's information is more valuable to the receiver when this sender's competitors have revealed less.

Theorem \ref{AoN} shows that all information can be transmitted from senders to the receiver under this condition and presents All-or-Nothing (AoN) strategies that support this equilibrium outcome. 
Each sender attracts attention proportional to the expected residual value of her information, i.e., assuming all competitors have revealed their information. 
We show that this amount constitutes a lower bound on attention for each sender. 
Therefore, this equilibrium is receiver-preferred -- information is transmitted in minimal time. 
In this equilibrium, each sender whose information has not been revealed offers an AoN probability that makes herself indifferent between being consulted now and 
 being consulted only after all competitors have revealed their information.
Since the receiver's information would include all competitors' components at that point, the  AoN probability and expected attention of each sender change over time as information is revealed. 

We discuss several implications that can directly be derived thanks to the tractable strategy and payoff characterisation in the equilibrium analysis.
Corollary \ref{Cor:submartingale} shows that the expected transmission speed in the AoN equilibrium increases over time. 
Corollary \ref{Cor:concentration} shows that concentrating a fixed amount of information on fewer senders hurts the receiver because each sender's attention is proportional to the residual value of their component. 
We then apply our general setup to the widely studied specification with Gaussian information and quadratic-loss preferences. 
Corollary \ref{Cor:symmetry} shows that the receiver prefers a fixed amount of conditionally independent information to be distributed evenly across senders.
Finally, Corollary \ref{Cor:correlation} shows that the receiver can benefit from correlation among information sources, even when this reduces the overall precision of the available information: correlation decreases the attention cost faster than the informational value.\footnote{The working paper version \cite{knoepfle2020dynamic} studies an additional application with endogenous sender-information. News sources determine the accuracy of their information in an investigation race in which they decide how long to investigate an issue before starting to report on it.}    


Finally, we consider the  competitive  limit as the number of senders grows large.\footnote{To keep the decision problem unchanged as the number of senders  increases, this part assumes each sender's component is independently and identically distributed conditional on some fixed payoff-relevant state. The general analysis allows for asymmetry as well as correlation among and payoff relevance of different senders' components.}    
Theorem \ref{AoN} already implies that the receiver learns the state exactly as the number of senders grows arbitrarily large, and that each sender's attention must go to zero.
Theorem \ref{prop:large-limit} shows additionally that the total amount of attention required by the receiver also goes to zero. In the limit, the receiver learns the exact state immediately.


	\paragraph{Related Literature.}
	
	This paper contributes to the literature on dynamic information design, first, by focusing on attention maximisation, and second, by proposing a tractable dynamic model with multiple competing senders who are partially informed. 
We contribute to the dynamic information acquisition literature by
endogenising the information processes chosen by senders.  

	This paper is related to the literature on \textbf{Bayesian persuasion}, based on \cite{kamenica2011bayesian}, in which a sender designs information to influence a receiver's behaviour. Senders in our model maximise attention and have no persuasion motive, that is, the action eventually taken by the receiver does not affect the senders' utilities.\footnote{Our results continue to hold when the senders' interests are aligned with the receiver regarding the decision problem but senders to not internalise the attention cost. For example, consultants and department heads want the CEO to take the best possible action but want to be the ones who are consulted most prior to it. For an analysis of competitive static persuasion where the attention and persuasion motives potentially conflict, see \cite{au2023attraction} and \cite{jain2020competitive}.} 
	
    \textbf{Dynamic information design} has received significant attention recently, as in \cite{kremer2014implementing, au2015dynamic, che2017recommender, renault2017optimal,  smolin2017dynamic, board2018competitive, ball2019dynamic,che2019listener,ely2019moving,guo2019interval,orlov2018persuading, escude2023slow, wu2023sequential, chen2024optimal, knoepfle2024dynamic, zhao2024contracting}, and others.
    The main contrasts to the current paper are the presence of the persuasion motive mentioned above and the focus on a single-sender.\footnote{With the exception of \cite{board2018competitive} who consider multiple sellers which are randomly matched with potential buyers in a search market and want to induce buyers to buy from them rather than continuing to search. For a cheap talk model that features multiple senders in a dynamic environment, see \cite{margaria2018dynamic}.}
    In more recent contributions, \cite{hebert2022engagement}, \cite{koh2022attention}, \cite{saeedi2024getting}, \cite{koh2024persuasion} study a problem related to the single-sender case of our model, where a monopolistic sender wants to maximise the attention from a receiver. 
    We discuss these papers together with the analysis of the monopoly case in Section \ref{sec:Monopoly}. 
  As in \cite{au2015dynamic} and \cite{che2019listener}, senders in the current paper commit to the information offered within a period but cannot commit across periods. The focus on the receiver's costly attention connects the current paper and \cite{che2019listener}, who examine optimal dynamic persuasion when the receiver has to pay an attention cost.

	Optimal \textbf{dynamic information acquisition} by a decision maker has been introduced to the economics literature by \cite{wald1947foundations}, where the decision maker decides when to stop observing an exogenous information process and take an action. Several papers enrich the decision maker's problem by allowing him to adjust the information intensity or to choose among exogenous processes, see \cite{moscarini2001optimal,mayskaya2017dynamic,che2019optimal,liang2019complementary,liang2019dynamically}, and others.\footnote{See also \cite{morris2017wald} and \cite{fudenberg2018speed} for recent developments on the Wald problem in different directions.} The decision maker in the current paper faces a related acquisition problem but chooses among experiments that are offered endogenously by the senders.
	\cite{zhong2019optimal} characterises the optimal information process designed by a decision maker with full flexibility facing a precision cost. He shows that the optimal policy consists of a Poisson process that leads to immediate action after arrival. 
	 The current paper contributes to the literature on optimal dynamic design by considering a related question from the opposite perspective. Senders choose flexibly how to provide their information over time to maximise the attention they attract from the receiver. The offer strategies presented in the current paper 
 imply a geometrically distributed arrival of all information from one sender. This is akin to a Poisson process in continuous time with fully revealing news and where the absence of arrival allows no inference (there is no belief drift). 

One interpretation of the setting in the current paper is to view senders as selling information to a receiver requiring a price in units of attention time. For a survey on \textbf{markets for information}, see \cite{bergemann2019markets}. \cite{bergemann2018design} study the optimal design and pricing of a menu of experiments to screen the receiver according to his willingness to pay. In the current paper, the receiver's willingness to pay is known. The monopolist can require attention proportional to the price charged by the monopolist in \cite{bergemann2018design}, who would only offer the fully revealing experiment if he knew the receiver's willingness to pay. 

The information-attention exchange studied in the current paper shares important features with \textbf{competitive pricing} for standard goods. 
First, the product `information' displays important externalities. This can hinder (efficient) trade already in static settings, see \cite{kelso1982job}, \cite{gul1999walrasian}, and \cite{tauman1997model}. We present an auxiliary static exchange akin to these papers in Section \ref{subsec:auxil} and then introduce a suitably extended version of their substitutes condition that enables the exchange of information. 
\cite{bergemann2006dynamic} study dynamic price competition with externalities -- the surplus of each purchase depends on the history of previous purchases. For the special case without inter-group externalities -- where the surplus generated by a trade with seller $i$ depends only on the number of previous trades with $i$ -- they show that a marginal contribution equilibrium exists and is efficient. In the current paper, information generally features inter-group externalities. Nevertheless, when they take the form of substitutes, we can construct an analogous equilibrium by considering each sender's expected marginal contribution.
Another related feature is the presence of capacity constraints: Each sender is initially endowed with information. This fixed endowment is a capacity constraint, relating the current paper to \cite{dudey1992dynamic,martinez2011dynamic}, and \cite{anton2014dynamic}. As information is assumed to be always worth one unit of attention for the receiver in the current paper, the capacity constraint eventually binds. Paralleling results in the above papers, this implies that each sender can extract positive surplus despite the competition.

\cite{gossner2021attention} study attention with a different focus. They show that drawing attention to one of several options unequivocally increases the likelihood of this option being chosen by an agent who uses a fixed threshold rule. This can be viewed as additional rationale to compete for attention.
	
\section{Model} \label{sec:model}

\subsection{Environment}

One receiver and $n$ senders $i \in N = \{1, \cdots, n\}$ interact in discrete rounds of communication, $t=0,1, \dots$. There is a state of the world $\boldsymbol\omega = \Paren{\omega_0, \omega_1, \cdots, \omega_n}$ in Polish space\footnote{A Polish space is a separable and completely metrisable space. This ensures the existence of the conditional probability measures used below.} $\Omega = \Omega_0 \times\Omega_1 \times  \cdots \times \Omega_n $. The state remains constant over time, and sender $i$ can reveal information about component $\omega_i$,  her `area of expertise'. For some applications it is convenient to include $\w_0$,  a payoff-relevant component for the receiver that cannot  be revealed by any sender $i$ directly but only through correlation with sender $i$'s component $\w_i$. Denote by $\boldsymbol{\w}_{-i} = (\w_j)_{j \in N \colon j \ne i}$ the vector of components without $\w_i$ and $\w_0$.  
 The state is distributed according to the common prior $\mu^0 \in \Delta(\Omega)$. Components may be correlated.

{The receiver} has to take an irreversible action $a$ from the closed space $A$ to maximise his utility $u(a, \boldsymbol\omega)$, where $u: A \times \Omega \rightarrow \mathbb{R} $ is continuous and bounded. For this, the receiver relies on information from the senders. In each round $t$, he can either pay the attention cost $c>0$ and consult one sender, or take an action with the information gathered so far. After the action is taken, the game ends.

{Senders} offer \textit{experiments} over their component. In line with the literature and to avoid signalling, assume that senders do not observe their component prior to revealing it through experiments.\footnote{For example, department heads within an organisation offer what data to gather or which study to commission within their area of expertise before actually processing the data.} An experiment is a conditional distribution over messages $m$ from the Polish space $M$. The message space $M$ is equal for all senders and rich enough to contain all information about $\boldsymbol{\omega}$, i.e.  $\Omega \subset M$.  Formally, at the beginning of round $t$, each sender $i$ simultaneously announces $ \lambda_{i,t}: \Omega_i \times \mathcal B(M) \rightarrow [0,1]$, a (regular) conditional probability such that $\lambda_{i,t}(\cdot, W)$ is measurable for all $W \in \mathcal B(M)$, and $\lambda_{i,t}(\omega_i, \cdot)$ is a probability measure given any realisation of component $\omega_i \in \Omega_i$. The set of possible experiments for sender $i$ is denoted by $\Lambda_i$.
Senders compete for attention. In each round that sender $i$ is consulted, she receives utility normalised to one.

\subsection{Strategies, payoffs, equilibrium concept}
First, nature draws state $\boldsymbol\omega$.  
At the beginning of each round $t \ge 0$, if the receiver has not taken an action previously, all senders simultaneously offer experiments $\lambda_{i,t}$. 

The receiver observes the offers and chooses $d_t \in \{0, 1,\cdots,n\}$, where $d_t=0$ encodes that he stops and $d_t = i \in\{1,\cdots,n\}$ means that he pays cost $c> 0$ and consults sender $i$.\footnote{The cost $c$ is interpreted as attention cost borne by the receiver, which is independent of which sender he consults. All results extend easily to the case in which the costs $c_i$ differ across senders.} When the receiver stops, he takes an action $a \in A$, the game ends, and payoffs realise.\footnote{For completeness, assume that $d_t = 0 \Rightarrow d_{t+1} = 0$.} Visiting sender $i$ implies that he observes $m_{i,t} \in M$ drawn from the distribution $\lambda_{i,t}$ and the game continues to the next round.\footnote{We assume that the receiver cannot wait without consulting any sender. If the receiver could wait at a cost, results remain unchanged. If the receiver could wait for free, under some decision-problems, this opens up the possibility of a threat-equilibrium in which senders offer no information and the receiver perpetually rejects/waits. We choose to rule out free waiting to eliminate this threat being leveraged to enforce equilibrium-payoff-profiles that are otherwise unattainable.
}

A public \textbf{history} of the game is 
\begin{align*}
&h^t = \left(\left( (\lambda_{i,1})_{i=1}^n, d_1, m_{d_1,1} \right) , \cdots, \left((\lambda_{i,t})_{i=1}^n, d_t, m_{d_t, t} \right) \right)& &\text{ for } t \ge 0,
\end{align*}
with initial history $h^{-1} = \emptyset$.
All players observe all past offers, the receiver's choices and the message of the chosen sender. Hence, all senders observe the information revealed by their competitors.\footnote{Given that experiments are offered publicly, it is plausible that the senders also learn from their competitors. For the Gaussian specification analysed at the end of Section \ref{subsec:AoN-eq}, this assumption is inconsequential.} Denote by $\mathcal H^t$ the set of round-$t$ histories. 

A pure \textbf{strategy} for the receiver is a collection of maps $\left( \sigma^R_t \right)_{t \ge 0}$ with  \begin{align*}
    \sigma^R_{t}: \mathcal H^{t-1} \times \left( \underset{i}{\times} \Lambda_i \right) \rightarrow \{0,1,\cdots, n\}. 
\end{align*}
Likewise, for each sender $i \in \{1,\cdots, n\}$, a pure strategy is a collection $\left( \sigma^i_t \right)_{t \ge 0}$ with \begin{align*}
    \sigma^i_{t}: \mathcal H^{t-1} \rightarrow \Lambda_i.
\end{align*}

The receiver's final \textbf{payoff} has two components. First, he gets utility $u(a,\boldsymbol\omega)$ when stopping with action $a$ if the state is $\boldsymbol\omega$. Second, there are attention costs that depend on how many times the receiver consults a sender before taking action. Each consultation costs the receiver attention cost $c>0$, so that his final payoff will be 
\begin{align*}
    u(a,\boldsymbol\omega) - c \cdot \sum_{t\ge 0}  \mathbbm{1}_{\{d_t \ne 0 \}}.
\end{align*}

Senders maximise the attention they attract. Senders do not care about the receiver's action $a$. Our results go through if senders share the receiver-preferences $u$ without the attention cost. Each consultation gives utility normalised to 1. Sender $i$'s final payoff is   
\begin{align*}
    \sum_{t\ge 0}  \mathbbm{1}_{\{d_t = i \}}.
\end{align*}


The \textbf{solution concept} is a Perfect Bayesian Equilibrium, with the additional requirement that beliefs are only updated with Bayes' rule according to the chosen experiment. This requirement ensures the `no-signalling-what-you-don't-know' property \citep[see][]{fudenberg1991perfect}, whereby offers do not reveal information the senders do not hold. It also implies that the experiment from an off-path offer, if accepted, would be interpreted correctly.

\subsection{Examples} \label{subsec:examples}
The above model nests many environments that have been used to study a variety of applications. Before moving to the analysis, we present three special cases of the general model. Applications for which these setups or a close variant have been used are mentioned in square brackets. 

\paragraph{Example 1: Gaussian information and quadratic loss.} \label{ex:1} [Global games: \cite{morris2002social, bergemann2013robust, angeletos2007efficient}. Social learning: \cite{vives1996social}. News markets: \cite{chen2019competition, galperti2018shared}]

The receiver wants to learn a state of the world $\w_0 \sim \mathcal N(0, 1/p_0)$ as precisely as possible. His utility from the action is $ u(a, \w_0) = - (a-\w_0)^2$, so that the expected stopping utility is minus the conditional variance $Var\left(\w_0 \vert \mu \right)$ given the current information.\footnote{The fact that $u$ is not bounded from below does not create problems here since the stopping utility at the prior is equal to $-\frac{1}{p_0} > - \infty $.} Each sender $i \in \{1, \cdots, n\}$ is endowed with a conditionally independent component $\omega_i \sim \mathcal N(\w_0, 1/p_i)$, where $p_i > 0$ is sender $i$'s \textit{precision level}.  \hfill \#

\paragraph{Example 2: Additive attributes.}
[Consumer search: \cite{wolinsky1986true, choi2018consumer,ke2019informational}. Advertising: \cite{anderson2009comparative,sun2011disclosing}]

Let the receiver be a consumer who considers buying one of two objects, $A = \{1,2\}$. The (net) utility of the two objects is $\boldsymbol\omega = (\omega_1, \omega_2) \in \mathbb{R}^2$, where each object's utility is determined by a common and an idiosyncratic attribute: 
\begin{align*}
    &\omega_i = Y + \gamma_i& &\text{ for } i \in \{1,2\}, &
\end{align*}
where $Y$ is the common component distributed according to $F$ on $[\underline Y, \bar Y]$ and $\gamma_i$ are distributed independently according to $G_i$ on $[\underline \gamma, \bar \gamma]$.
Each sender holds information about one of the options in the form of a noisy component of the total utility but is unable to distinguish between the common and the idiosyncratic component. \hfill \#

\paragraph{Example 3: Hypothesis testing.}
[\cite{wald1947foundations}. Value and demand for information: \cite{moscarini2002law}]

There are two possible hypotheses $\w_0 \in {H_0,H_1}$ and the receiver must choose one of them, $A= \{0,1\}$. With the utility function $u(1,H_0) =-\alpha $,  $u(0,H_1) =-\beta$, and  $u(0,H_0) = u(1,H_1) =  0$, the parameters $\a,\b >0$ denote the cost of a type-I and type-II error, respectively.  \hfill \#



\section{Updating and the value of information} \label{sec:Updating}

In this Section we derive expressions for the value of each sender's information that facilitates a tractable exposition of the remainder of the analysis.
 The receiver's \textit{stopping utility} with belief $\mu$, i.e. the expected utility from the optimal action, given that he currently holds posterior $\mu$, is 
\begin{align*}
   U(\mu) \equiv  \max_{a\in A} \E_{\boldsymbol\omega \sim \mu } \left[\; u(a,\boldsymbol\omega) \; \right].
\end{align*}
If, while holding belief $\mu$, the receiver expects to learn the state precisely and take the optimal action in each state, his expected utility is 
\begin{align*}
  \bar U(\mu) \equiv  \E_{\boldsymbol\omega \sim \mu } \left[ \;  \max_{a\in A} \ u(a,\boldsymbol\omega) \; \right].
\end{align*}
Clearly $\bar U$ is an upper bound of $U$; the two coincide at beliefs that reveal all payoff-relevant information. 

At the end of round $t$, the belief $\mu^t$ is updated to $\mu^{t+1}$ by Bayes' rule after observing message $m_t$ resulting from the selected experiment $\lambda_t$. Denote the \textit{updating rule} by $\mu'$ such that \begin{align*}
    \mu^{t+1}=\mu'(\mu^t, m).
\end{align*}
This notation suppresses the chosen experiment in the updating rule. Appendix $\ref{AppProofs}$ presents the formal definition of the updating rule and shows that it is well defined.

Define the \textit{value} of offer $\lambda_i$ at current belief $\mu$ as as the expected difference between the stopping utilities with and without the additional information from $\lambda_i$:
\begin{align*}
    v\left( \lambda_i \vert \mu \right) \equiv \E_{\omega_i \sim \mu }\Big[  \E_{m \sim \lambda_i(\omega_i,\cdot)}\left[ U\left(\mu'(\mu,m)\right) \right]\Big]  - U(\mu) . 
\end{align*}
Note that $v \ge 0 $ always because $U$ is convex. Value $v$ may be lower than the increase in the receiver's expected equilibrium payoff because the receiver may not stop after experiment $\lambda_i$. That is, $v$ ignores the receiver's option value. We use this more `mechanical' notion to define a value independently of the players' strategies.
 The value of learning a sender's component $\w_i$ exactly is denoted by 
\begin{align*}
\bar v (\tilde \w_i  \vert \mu  ) \equiv  v\left( \delta_{\{\omega_i\}}\vert \mu \right).
\end{align*} 
 Here, the experiment that reveals $i$'s component precisely is denoted by Dirac measure $\lambda_i(\omega_i, \cdot) = \delta_{\{\omega_i\}}(\cdot)$. 
 Write simply $\bar v (\tilde \w_i )$ without the second argument when $\mu = \mu^0$, that is when nothing has been learnt.  

The following class of histories plays a prominent role in the analysis. Suppose the receiver has learned the realisation of components $\boldsymbol{\w_S}$ for some subset of senders $S\subset N$ but has not learned any information directly from senders $i \notin  S$. 
We slightly abuse notation and denote the corresponding posterior by $\mu'( \boldsymbol{\omega}_{S})$, and the value of learning $i$'s component at this history by 
\begin{align*}
\bar v (\tilde \w_i  \vert \boldsymbol{\w}_S ) \equiv  v\Big( \delta_{\{\omega_i\}}\vert \mu'(\boldsymbol{\omega}_{S}) \Big).
\end{align*}





\section{Single sender}\label{sec:Monopoly}

Consider the case of a monopolistic sender $n=1$. Without loss, let the sender have access to all payoff relevant information, for example setting $\w_0 = \w_1 $.
It is necessary for any information transmission to be feasible that attention can be split finely enough to make the sender's information worth at least one unit of attention:
\begin{Ass}\label{Ass1}
\begin{align*}
    c< \bar v \left(\tilde \w_1\right). 
\end{align*}
\end{Ass}

What is the maximal expected attention cost the receiver is willing to pay for the sender's information? It is equal to $\bar v \left(\tilde \w_1\right) = \bar U(\m^0) - U(\mu^0)$, the difference between the expected stopping utility with all information and the stopping utility with no information. As each visit requires a cost of $c$, the maximal expected number of visits the sender can attract is $ \bar v \left(\tilde \w_1 \right)/c$.

If the sender had intertemporal commitment, the simplest strategy to implement this outcome would first require ${ \bar v \left(\tilde \w_1  \right)}/{c}-1$ visits from the receiver at which no information is revealed, and then reveal all information at the last visit.\footnote{This illustration argument neglects non-divisibilities that make this potential strategy suboptimal as it could only achieve integer amounts of visits.} However, the sender lacks the intertemporal commitment to credibly promise all information in the last round -- she may, for example, repeat the round-0 strategy.

A simple sender strategy to overcome the non-commitment issue and to deal with potential integer problems is to offer revealing $\omega_1$ with some probability $\lambda \in [0,1]$ and revealing no information with probability $1-\lambda$.\footnote{To keep the expressions simpler, we abuse notation by letting $\lambda$ denote a probability in [0,1], while $\lambda$ generally denotes distributions over messages. Formally, the AoN experiment is represented by the conditional distribution $\lambda(\omega_1, \cdot) =  \lambda \delta_{\{\omega_1\}} + (1-\lambda)\delta_{\{m\}}$ for an arbitrary message $m \in M \backslash \Omega_1 $ that conveys no information.} Offers in this class are denoted \textit{All-or-Nothing} (AoN) offers.

The following result shows that an equilibrium in AoN strategies generally exists in the monopoly game. Furthermore, the payoffs are the same in any equilibrium.
\begin{Thm} \label{thm:monopoly}
Let $n=1$. In any equilibrium of the monopoly game 
the expected payoffs are $
{ \bar v(\tilde \w_1)}/{c}$ for the sender and $U(\mu^0 )$ for the receiver.
There is an AoN equilibrium in which, in each round, the sender makes an AoN offer with revelation probability   
\begin{align*}
     \lambda^\text{M} = \frac{c}{\bar v(\tilde \w_1 )},
\end{align*}
and the receiver accepts every round until $\omega_1$ is revealed.
\end{Thm}

Note that Assumption \ref{Ass1} ensures that $  \lambda^\text{M} <1$.
If the sender's strategy prescribes AoN offers until all information is transmitted, the receiver's continuation payoff in the event of no revelation (which happens with probability $1-\lambda^\text{M}$) remains at his initial payoff. Hence, the AoN probability $ \lambda^\text{M}$ that makes the receiver indifferent between taking action immediately and accepting the offer satisfies
\begin{align*}
  U(\mu^0) = - c +  \lambda^\text{M}  \ \bar U(\mu^0 ) + (1- \lambda^\text{M})\ U(\mu^0).
\end{align*}
Observing the experiment costs $c$. With probability $ \lambda^\text{M}$, the receiver learns $\omega_1$ and stops with full information, which gives expected utility of $\bar U(\mu^0 )$.
With probability $1- \lambda^\text{M}$, the receiver learns no information, which gives utility $U(\mu^0)$ as the sender will keep him indifferent in the following round again. 

The offer $ \lambda^\text{M}$ is accepted by the receiver in every round until the information is eventually revealed. The number of rounds until revelation follows a geometric distribution with parameter $ \lambda^\text{M}$, so that the expected number of rounds is ${1}/{\lambda^\text{M}}$.
As the receiver is indifferent between accepting and stopping in every round, it should not be a surprise that solving the above indifference condition for $\lambda^\text{M}$ gives
\begin{align*}
    \lambda^\text{M} =\frac{c}{\bar U(\mu^0) - U(\mu^0)} = \frac{c}{\bar v(\tilde \w_1 )}.
\end{align*}
The expected attention is precisely the upper bound the receiver is willing to spend.

Theorem \ref{thm:monopoly} implies that the sender would not benefit from intertemporal commitment. Another implication of Theorem \ref{thm:monopoly} is that the receiver's equilibrium continuation payoff in each period $t$ has to equal his stopping payoff $U\Paren{\mu^t}$.  
While equilibrium payoffs are unique, depending on the information structure and decision problem, there may be  other strategies that resolve the sender's lack of commitment and attract the maximal amount of attention.
 The AoN process describes a particularly simple strategy that works for all environments described above.
The AoN process in the monopoly is stationary and the belief remains constant until it jumps to certainty in the last round of the game. An attractive feature of stationarity is that the sender does not need to observe the receiver's precise arrival time, or the times at which the receiver consulted the sender in order to know what information to offer next. 

The result above is robust to common discounting; the analysis remains almost unchanged when the receiver and the sender share the same discount factor.
If the sender is more patient than the receiver, then the uniquely optimal transmission process is an AoN process starting from the second period  as shown by \citealp{koh2022attention}. This process maximises the variance of the revelation time to exploit the receiver's different risk preferences over time lotteries.\footnote{In a previous version of this paper we conjectured that the AoN process in Theorem \ref{thm:monopoly}, where the belief remains equal to the prior until the geometric revelation time, was (uniquely) optimal in this case.  \cite{koh2022attention} and \cite{saeedi2024getting} show for the binary-state case that this holds only for some prior beliefs. When the prior does not maximise the value of information $\bar v (\tilde \w_t \lvert \mu^0)$, then it is optimal to initially reveal one state with relatively higher probability so that the receiver's belief moves to the maximiser of $\bar v$ and only then uses the AoN strategy  above.}

\subsection{Gradual information transmission vs necessity of jumps}

We present two examples based on the specification in Example \hyperref[ex:1]{1} to illustrate the AoN equilibrium from Theorem \ref{thm:monopoly}. We demonstrate that an alternative, fully gradual, transmission process is feasible in a decision problem with a continuum of actions, whereas a finite action space requires information to arrive in jumps. 
To abstract from a `mechanical' driver of jumps given by the discrete rounds with a strictly positive cost per unit of attention, the examples below consider the continuous-time analogue of the model where the length of each period $\de t$ approaches $0$ and the attention cost per instant is $c \de t$.\footnote{The formal continuous-time model is easily specified in the monopoly case, where we can equivalently let the sender choose a belief martingale (subject to sequential rationality). The general model is set in discrete time to avoid complications in the multiple-sender case where each sender offers an arbitrary Blackwell experiment about only her own component at each instant.}


 \paragraph{Example 1a: Gradual transmission.} \label{ex:1a}
 Suppose $\w_1 \sim \mathcal N(0, 1/p)$ and $ u(a, \w_1) = - (a-\w_1)^2$, so that $U(\mu) = - \Var\left(\w_1 \vert \mu \right)$. Theorem \ref{thm:monopoly} tells us immediately that the sender extracts  ${ \bar v(\tilde \w_1 )}/{c} = 1/(pc)$ units of attention in any equilibrium. 
This can be implemented with (the continuous-time analogue of) an AoN process: a fully revealing signal arriving at Poisson rate  $\l = p c$. 

Alternatively, in this setting with a continuum of actions, the information can be transmitted gradually from time $t=0$ to fixed final time $T^* \equiv 1/( p c)$. 
For example, let the sender offer to reveal, as long as the receiver pays attention, the realisation of the process 
 \begin{align*}
    X_t =  \frac{t}{T^*} \cdot \w_1  + \sqrt{c} \Paren{W_t - \frac{t}{T^*} \cdot W_{T^*}} ,
\end{align*}
where $(W_t)_{t\ge 0}$ is a standard Brownian Motion. Process $X$ presents a Brownian bridge from $X_0 = 0 $ to $X_{T^*}= \w_1$.
It is easily verified that the mean-squared error of the prediction of $\w_1$ based on $X_t$ is equal to \ $c t - 1/p$. 
Thus, the receiver's stopping utility satisfies $U(\mu^t) = -1/p +c t$ and he is indifferent. His loss is reduced linearly and deterministically until he knows the exact state $\omega_1$ at time $T^*= 1/( p c)$. \hfill \#

The next example changes the agent's decision problem to illustrate that gradual revelation is no longer possible with finite actions. 
Note that full surplus extraction in Theorem \ref{thm:monopoly} implies that the receiver's continuation utility in any equilibrium is equal to $U(\mu^t)$ at all $t$. When the sender lacks intertemporal commitment, the receiver is indifferent between continuing and stopping at all times.  
Further, if the action space $A$ is finite, then $U$ is a piece-wise affine function of the current belief. Linear segments correspond to beliefs at which the same action is optimal. Naturally, information that does not change the optimal action with positive probability, is not valuable.  The next example shows that this `non-concavity in the value of information' \citep[see][]{radner1984nonconcavity} and the lack of intertemporal commitment jointly imply that information must arrive in jumps. 

 
\paragraph{Example 1b: Necessity of jumps.} \label{ex:1b}
Let  $\w_1$ be as above, $A= \{-1, +1\}$, and the utility $u(a, \w_1) = a \w_1$. The receiver chooses $a=1$ iff the expected state is positive. The stopping utility is $U(\mu) = \Abs{ \E_{\w_1 \sim \mu}\Brac{\w_1}}$ and the expected utility with full information is $\bar U(\mu) =  \E_{\w_1 \sim \mu}\Brac{\Abs{\w_1}}$. Theorem \ref{thm:monopoly} implies that the sender extracts surplus ${ \bar v(\tilde \w_1 )}/{c} = \sqrt{2/(\pi p )}/c$. 
This can be implemented with a fully revealing signal arriving at Poisson rate $\l =c / \sqrt{2/(\pi p )}$.

By contrast to Example 1a above, gradual information transmission is not feasible in any equilibrium. 
For any belief $\mu^t$, denote the expected state by $Y_t =  \E_{\w_1 \sim \mu^t}\Brac{\w_1}$. Then we have $U(\mu^t) = \Abs{Y_t}$. 
For any c\`{a}dl\`{a}g  
 process $Y$, the Meyer-It\^{o} Formula \citep[see, for example, ][Theorem 70]{protter2005stochastic} gives
\begin{align} \label{eq:Ito}
    \Abs{Y_t} - \Abs{Y_0} = \int_{0+}^t \sgn(Y_{s-}) \de Y_{s} + 2 \frac{1}{2} L_t^{0} + \sum_{0<s \le t} {\big (} \Abs{Y_s} - \Abs{Y_{s-}} - \sgn(Y_{s-})\Delta Y_s {\big )} .
\end{align}
 The integral term is the usual first-derivative part in the standard It\^{o}'s formula for $C^2$-functions and continuous processes; $sgn(Y_{s-})$ is the first derivative of the absolute value. The second term corresponds to the second-derivative part. Note that the second derivative of $\Abs{Y}$ is 0 whenever $Y\ne 0$. The term $1/2 L_t^{0}$ is the local (occupation) time of process $Y$ at position $0$ until time $t$, and the factor 2 can informally be thought of as the second derivative at 0.\footnote{For a formal treatment of local times for semimartingales with jumps, see \citet[Section IV.7]{protter2005stochastic}.}  Finally, the sum in the last term considers jumps in $Y$. 
 
 Taking expectations to determine the value of information, we see that purely gradual transmission cannot be an equilibrium.  Receiver-indifference requires $ \E\Brac{\Abs{Y_t} - \Abs{Y_0} }= c t$ for all $t$.  Since $Y$ is a martingale, the first term on the right-hand side of \eqref{eq:Ito} vanishes and $  \E\Brac{\Abs{Y_t} - \Abs{Y_0} }= \E\Brac{ L_t^{0} +  \sum_{0<s \le t} {\big (} \Abs{Y_s} - \Abs{Y_{s-}} - \sgn(Y_{s-})\Delta Y_s {\big )}}$. This illustrates that any valuable change in belief must move the expectation $Y_t$ from one side of the indifference value $0$ to the other side. In particular, when $Y_{s-}$ and $Y_{s}$ are both negative or both positive, the entire term in the sum is 0 because then $\Abs{Y_s} - \Abs{Y_{s-}} = \sgn(Y_{s-})\Delta Y_s $. Thus, information transmission in equilibrium requires a jump-component at all times to ensure that the optimal action changes with high enough probability.
 \hfill \#

Attention or engagement maximisation by a monopolistic sender has been the focus of recent contributions (see \citealp{hebert2022engagement}; \citealp{koh2022attention}; \citealp{saeedi2024getting}; and \citealp{koh2024persuasion}).
 \cite{hebert2022engagement} also find that jumps in the information process are necessary for sender optimality. The reasons for jumps are different. In \cite{hebert2022engagement} jumps are necessary even with sender-commitment as the sender forces more information upon the receiver than the receiver would acquire at the attention cost. In the above example, jumps are necessary because the sender must deliver positive value to commit herself to deliver information fast enough in the future. 
A natural question in view of Example 1b 
 is whether the lack of intertemporal commitment necessarily introduces stochasticity in the receiver's stopping time when there are finitely many actions. \cite{koh2022attention} show that the answer is no: a single sender without commitment can implement any (commitment-) optimal distribution of stopping times. 
 \cite{koh2022attention} construct an equilibrium in which the sender `commits' herself to reveal information in a way that leads to the same distribution of exit times. However, Example 1b above indicates that while the stopping time may be deterministic, the information yields surprised with positive probability at all times.

\section{Multiple Senders}\label{sec:multiple senders}

Consider the general case with $n\ge 2$ senders. To make sure senders can always attract at least one unit of attention, we make an analogous assumption to Assumption \ref{Ass1} for the multiple-sender case. 
\begin{Ass}\label{Ass2}
For all $i \in \{1,\cdots,n\}$ and for all $\boldsymbol{\omega}_{-i}$ in the support of $\mu^0$:
\begin{align}
   c<  \bar v \left(\tilde \w_i \ \vert \ \boldsymbol{\w}_{-i}\right).  
\end{align}
\end{Ass}
This condition ensures that for any realisation of the other senders' components, sender $i$'s information is sufficiently valuable to attract at least one visit. Assumption \ref{Ass2} implies that no sender has perfect information about the payoff-relevant state. Due to the competition in our game, if a sender had perfect information, all other senders would offer all information immediately. With two or more perfectly informed senders, the receiver would become perfectly informed after one visit. 
Assumption \ref{Ass2} makes sure that there will be no exclusion of any sender on the `extensive' margin in that no sender becomes incapable of attracting at least one unit of attention. 
Note that the corresponding effect of competition on the `intensive' margin -- senders extract less attention because of the presence of competitors -- will play an important role in the analysis below.\footnote{\label{fn:exclusion}We rule out exclusion for tractability. There are two simple ways to amend any game that does not satisfy Assumption \ref{Ass2} to rule out exclusion. The first is to enlarge the action space of each sender and allow her to post with each experiment, an amount of attention $\tilde c_{i,t} \in [0,c]$ required from the receiver to get the information from the posted experiment, and which will deliver the sender a benefit of $\tilde c_{i,t}/c \in [0,1]$ in that round. With this modification, $c$ determines an upper bound of attention that can be garnered in each round. While still imposing gradual revelation and sequentiality, senders can always reveal even more gradually. Second, we can let each receiver's action set include an additional and relatively `unimportant' choice from some $A'_i$ corresponding to each sender $i$ and add to each sender $i$'s component independent information that increases the value from that action. Choosing $c$ small enough, i.e. considering small rounds of attention, then always satisfies the assumption. To keep the exposition clean, we do not include either of these modifications explicitly, but impose the mild Assumption \ref{Ass2} throughout. 
} 



Information transmission in the dynamic game is potentially thwarted by two innate features of the attention-information exchange. 
First, attention can be collected only over time, requiring an element of sequentiality in the transmission process. In contrast to monetary transfers, the receiver cannot  spend a large amount of attention at one instant in exchange for a fixed amount of immediate information. Instead, in order to extract a significant amount of attention, each sender needs to offer information piece-wise, facing competition by other senders in each round.
Second, information is a complex good that can be transferred from the senders to the receiver through potentially complex transmission processes. Every round, each sender chooses from an infinite-dimensional action space of Blackwell experiments, in contrast to the one-dimensional quantity or quality choice for more traditional goods.   

Additionally, the informational components from different senders naturally display externalities. The
information from one sender changes the receiver's need for further information and his probability assessment of the unknown pieces of information.  
Before presenting the AoN equilibrium of the dynamic game, the following subsection presents a simple auxiliary market economy to abstract from
the two potential complications above and allow us to focus on the role of externalities. 
Readers familiar with pricing and competition with consumption externalities (see \citealp{gul1999walrasian}; \citealp{kelso1982job}; \citealp{tauman1997model}) may skip the following subsection and jump directly to the equilibrium analysis of the original game in Subsection \ref{subsec:AoN-eq}. There, we establish a condition on the externalities under which the competitive equilibrium outcome of the auxiliary economy can be attained as an equilibrium of our original game.

\subsection{An auxiliary exchange economy} \label{subsec:auxil}
This benchmark considers a one-shot exchange with transfers. Let the receiver's gross utility when choosing consumption set $S \in 2^N$ be defined by  
$$f(S) =  \E_{\tilde{\boldsymbol{\w}}_S \sim \mu_0 }\left[   U\left(\mu'(\tilde{\boldsymbol{\w}}_S)\right) \right].$$
That is, $f(S)$ is the ex-ante expected increase in the stopping utility $U$ from learning the realisation of $\w_i$ for each $i\in S$ and learning nothing directly from senders $j \notin S$.
Given a price vector $\boldsymbol{p} = (p_1, \dots,p_n) \in \mathbb{R}^n_+$ and a consumption set $S\in 2^N$, the net utility of the receiver is $f(S) - \sum_{i\in S} p_i$, and the utility of sender $i$ is $p_i$ if $i\in S$ and 0 otherwise.  

Deriving $f$ from the primitives of our model per the definition above, with the convention $f(\emptyset) =0$, we can use existing results (see \citealp{gul1999walrasian} and \citealp{tauman1997model}) to analyse the equilibria of this exchange economy. \cite{gul1999walrasian} analyse a competitive setting in which firms are price takers. By contrast, in \cite{tauman1997model} firms set (commit to) prices simultaneously and the consumer chooses which bundle to purchase. 
In our model, the receiver and the senders interact repeatedly and senders cannot commit to future offers, so the price-setting power of our senders lies between these two extremes.

We discuss a condition that different components form \textit{substitutes}, and will impose a suitably extended form on our information exchange in the following subsection.   
\cite{kelso1982job} introduce a condition on preferences, \textit{Gross Substitutes} (GS), and show that it is sufficient for equilibrium existence in their more general setup.\footnote{They allow for multiple players on both sides and for more general preferences for money. \citet[p. 103]{gul1999walrasian} show that `in a sense GS is a  ``necessary'' condition for existence' in that it defines the largest class of preferences which contains all unit-demand preferences for which equilibrium  existence is guaranteed.}
The GS condition in \cite{kelso1982job} and \cite{gul1999walrasian} is expressed in terms of the implied demand correspondence. According to their definition, goods are gross substitutes if an increase in the prices of any subset of other goods weakly increases the demand of each good whose price remains constant. This was subsequently  shown to be equivalent to $M^\natural$-concavity\footnote{$M^\natural$ stands for M-natural.} of the valuation function \citep{fujishige2003note}.
\begin{Def}  The valuation function $f$ is $M^\natural$-concave if for all $S,T \subseteq N$ and $s \in S \backslash T$  
\begin{align*}
    f(S)+f(T) \le \max\left\{ f(S-s) + f(T+s) \ ; \ \max_{t\in T\backslash S}\{ f(S-s+t) + f(T+s-t)  \} \right\}.
\end{align*}
\end{Def}
Under this condition, the equilibrium consumption sets in \cite{gul1999walrasian} and \cite{tauman1997model} coincide and there is a single price vector that satisfies the equilibrium conditions in both worlds with price-taking and price-setting firms.\footnote{The condition in \cite{tauman1997model} takes the weaker form of decreasing differences in $f$. This is sufficient in their setting with a single consumer who makes a once and for all purchase decision.} 
The equilibrium consumption bundle is the entire set $N$ and the price paid to each firm $i$ is $p_i =f(N)- f(N-i)$. 

By definition of $f$, the equilibrium price paid to firm $i$ is $f(N)-f(N-i) = \E_{\boldsymbol{\w}_{-i}\sim \mu^0} \Brac{ \bar v(\tilde \w_i \vert \boldsymbol{\w}_{-i}) }$, i.e. the receiver's expected value of learning $\w_i$ after the information from all firms different from $i$ has been learnt. Each firm is paid her marginal contribution to the receiver's utility.\footnote{This outcome is also in the core as defined in \cite{gul1999walrasian}. This would also be the case in \cite{tauman1997model} if the definition  takes the consumer into account. The core defined in \cite{tauman1997model} excludes the consumer and is therefore empty when goods are substitutes.}

Now we show that the above outcome of the one-shot exchange with transfers can be implemented in our original game where attention has to be collected gradually and sequentially across senders, with senders reacting dynamically to what information has previously been revealed. 
We identify a substitutes condition on the receiver's preferences, expressed in terms of the value function $\bar v$, which captures the additional complexity when the different components represent information.   
We then show that under this substitutes condition there is an equilibrium in which senders use time-varying AoN offers such that all information is revealed to the receiver and the expected attention of each sender is equal to the marginal-contribution price identified above.

\subsection{Equilibrium analysis of the original game} \label{subsec:AoN-eq}

 We now construct an equilibrium of the original game that leads to the same expected payoff as the equilibrium in the auxiliary exchange. 
 
To see how externalities may impede information transmission in our dynamic game, consider the following example (a detailed argument is presented in Appendix \ref{subsec:compl}). Suppose there are two senders. Each sender's component is an independent, fair coin flip. The receiver has to guess whether the two coins match or not. For this decision problem, the components are perfect complements. Each component is valuable only in conjunction with the other. When senders cannot commit across rounds, complements can cause a hold-up problem: after one sender has revealed her information, the following sender would require maximal attention, keeping the receiver at the current stopping utility with one component only. Anticipating this, the receiver is not willing to spend any attention on the first component given that it delivers no value on its own.

The following condition on the externalities rules out this sort of problems and allows for full information transmission. The likelihood-ratio condition below implies that posterior belief $\m$ contains no direct information from sender $i$.  
\begin{Def}
Senders' components are substitutes if the following inequality holds for all senders $i$ and all beliefs $\mu$ such that 
$\frac{\mu(\omega_i, W_{-i})}{\mu(\w'_i, W_{-i})} = \frac{\mu^0(\omega_i, W_{-i})}{\mu^0(\w'_i, W_{-i})}$ for all $\w_i, \w_i'$ and all $W_{-i} \in \BB(\W_{-i})$.
\begin{align}\label{eq:subst}
    \bar v\left(\tilde \w_i \ \vert \  \mu \right) \ge \E_{\boldsymbol{\omega}_{-i} \sim \mu} \left[  \bar v\left(\tilde \w_i \ \vert \  \boldsymbol{\w}_{-i} \right)  \right].\tag{SU}
\end{align}
\end{Def}

Signals are substitutes if the  value of learning $\omega_i$ is greater than the expected value of learning $\omega_i$ after having learnt the components of all competitors.
That sender $i$'s information is more valuable the less is known from her competitors is consistent with many applications. This is especially the case when senders report on a single issue or, as in Example 2, when the components contain information about a common component that affects all options. 
The substitutes condition is related to \cite{borgers2013signals}, who introduce a notion of substitutes for pairs of components. 
Condition \eqref{eq:subst} considers a fixed utility function at varying posterior beliefs whereas \cite{borgers2013signals} require the inequality to hold for all decision problems but only ex-ante, i.e. at $\mu = \mu^0$. It can be shown that the condition in \cite{borgers2013signals} implies \eqref{eq:subst}.

The following result confirms that the competitive outcome from Subsection \ref{subsec:auxil} can be attained in equilibrium when the senders' components are substitutes. This equilibrium attains full information transmission in the shortest possible time among all equilibria, making it receiver-preferred. 

\begin{Thm} \label{AoN}
If senders' components are substitutes, there is an equilibrium with the following strategies. At belief $\mu$, senders whose information has not been revealed offer AoN experiments with probability \begin{align} \label{eq:LamStar}
\lambda_i^*(\mu) =  \frac{c}{\E_{\boldsymbol{\omega}_{-i}\sim \mu} \left[\bar v\left(\tilde \w_i \ \vert \ \boldsymbol{\w}_{-i}\right) \right]}.
\end{align} 
The receiver is indifferent between visiting any of the senders whose information has not been revealed and visits them in arbitrary order until all information is transmitted. 
 The expected payoffs are \begin{align*}
 \begin{cases}   \E_{\boldsymbol{\w}_{-i}\sim \mu ^0} \left[ { \bar v \left(\tilde \w_i \lvert \boldsymbol{\w}_{-i} \right)} \right] \Big / {c} &\text{ for sender } i \vspace*{1em}  \\
  \E_{\boldsymbol{\w}_{N}\sim \mu ^0} \left[ U (\mu(\boldsymbol{\w}_N)) \right] - \underset{i\in N}{\sum}\E_{\boldsymbol{\w}_{-i}\sim \mu ^0}\left[\bar v \left(\tilde \w_i \lvert \boldsymbol{\w}_{-i} \right) \right]  &\text{ for the receiver.}
 \end{cases}\end{align*}
 This equilibrium is receiver-preferred.
\end{Thm}
Theorem \ref{AoN} provides for a simple computation of equilibrium payoffs. It is sufficient to compute the expected residual value of each sender's information. 
The sender payoffs reveal that the receiver pays attention $ \E_{\boldsymbol{\w}_{-i}\sim \mu ^0} \left[ { \bar v \left(\tilde \w_i \lvert \boldsymbol{\w}_{-i} \right)} \right]$ to each sender $i$, the same as the price paid to $i$ in the auxiliary exchange above. 
The traffic each source can attract is determined by the specialised pieces of information which are not in common with other sources directly or through correlation. 

The receiver-preferred equilibrium is Pareto-efficient among all feasible outcomes as any additional benefit to a sender has to stem from additional attention-costs borne by the receiver. Since information has no instrumental value for the senders, the receiver's payoff best captures the tradeoff between the amount and the speed of information transmission. Welfare comparisons depend on our normalisation of the value each visit gives to the sender. If the attention cost $c$ exceeds the senders' value derived from a visit, the receiver-preferred equilibrium is also welfare-maximising.

While AoN offers are stylised, the information transmission displays many aspects observed when searching for information online, such as tutorials or cooking recipes. The title informs the potential visitor what he will learn prior to visiting the site. 
However it is often unknown how long it takes to learn the promised information. In text, because the information is hidden within redundant paragraphs; in videos, because it is unclear when the informative piece is located. While reading or watching the content from one source, the visitor may change to another source at any time. 

After providing intuition for the equilibrium construction, 
the rest of this section derives several implications of the result concerning the dynamics of information transmission and the optimal distribution of information across senders. 

The Theorem 2 shows, for websites, that the traffic they can attract is determined by the
specialised pieces of information not in common with other sources.

The first step of the proof shows that if sender $i$ has not revealed any information by round $t$, then $\E_{\boldsymbol{\w}_{-i}\sim \mu^t} [ \bar v(\tilde \w_i \lvert \boldsymbol{\w}_{-i})]/c$ constitutes a lower bound for sender $i$'s continuation payoff.  This is intuitive as $\bar v(\tilde \w_i \lvert \boldsymbol{\w}_{-i})/c$ is the payoff sender $i$ gets if her competitors reveal $\boldsymbol{\w}_{-i}$ first, and then sender $i$ becomes the monopolist on the receiver's `residual' demand. 
By \eqref{eq:subst}, sender $i$'s information is more valuable when competitors have revealed less, so sender $i$ can secure this payoff for any strategies of her competitors. 

If players follow the strategies in Theorem \ref{AoN}, then sender $i$'s continuation payoff at $t$ is equal to $\E_{\boldsymbol{\w}_{-i}\sim \mu^t} [ \bar v(\tilde \w_i \lvert \boldsymbol{\w}_{-i})]/c =1/ \l^*_i(\mu^t)$. 
This is obvious if, starting from $t$, the receiver repeatedly consults only sender $i$ until her AoN experiment reveals $\w_i$. The number of visits $i$ enjoys is  geometrically distributed with parameter $\l^*_i(\mu^t)$. 
If, at time $t$, the receiver consults a different sender, then sender $i$'s upcoming AoN probability $\l^*_i(\mu^{t+1})$ may be higher or lower than $\l^*_i(\mu^t)$, depending on the information revealed by the chosen competitor. However, sender $i$'s time-$t$ expected payoff is the same because $\E_{\boldsymbol{\w}_{-i}\sim \mu^t} [ \bar v(\tilde \w_i \lvert \boldsymbol{\w}_{-i})]/c $ is a martingale. Put differently, when facing competition, the equilibrium offer strategy makes sender $i$ indifferent between being accepted or rejected.  

Finally, we confirm that there are no profitable deviations. 
The $\lambda_i^*$ are constructed to make the receiver indifferent between learning $\boldsymbol{\w}_{N}$ and learning only $\boldsymbol{\w}_{-i}$ but saving the attention cost  ${1}/ \lambda_{i}^*$ at sender $i$
Suppose $i$ deviates to a strategy that results in an expected payoff strictly higher than ${1} / \lambda_{i}^*(\mu^t)$. 
Given that the competitors follow the AoN equilibrium strategies, the receiver is strictly better off rejecting $i$'s offer forever even if $i$'s deviation strategy would lead to $\w_i$  being revealed fully.

\begin{Rem}
The equilibrium strategies in Theorem \ref{AoN} are Markov on the equilibrium path. Senders' actions are fully determined by the state, $\mu$. The receiver's actions are fully determined by the state and the offers made in the current round. Due to the complex action spaces of the senders, we cannot establish the existence of Markov perfect equilibria relying on results in the literature and chose a constructive approach.   
\end{Rem} 

Next, we leverage the tractable characterisation in Theorem \ref{AoN} to explore equilibrium dynamics and derive comparative statics. 

First, informative realisations become more likely over time. 
\begin{Cor}[Transmission rate dynamics] \label{Cor:submartingale}
Let $\l^*_{i}(\mu^t)_{t\ge 0}$ be the offer process of sender $i$ in an AoN equilibrium. 
While sender $i$'s component remains unknown $\E^t \Brac{\l^*_{i}(\mu^{t+1})} \ge  \l^*_{i}(\mu^t) $.
\end{Cor} 
The probability of revelation for sender $i$, which depends on the revelations of previous senders through the change in $i$'s residual value, increases in expectation. This follows immediately from the construction of $\l^*_{i,t}$. Since $(  \l^*_{i}(\mu^t) )^{-1}$ is a martingale, $  \l^*_{i}(\mu^t) $ is a submartingale. 

Second, concentrating the same amount of information on fewer sources hurts the receiver.
\begin{Cor}[Concentration of information]\label{Cor:concentration}
Let $n\ge 2$ and suppose senders' components are substitutes. 
If senders $1$ and $2$ are replaced by a single sender with component $\omega_{1,2} = (\w_1, \w_2)$, then the receiver's expected payoff in the AoN Equilibrium decreases. 
 \end{Cor}
By Theorem \ref{AoN} 
 the receiver's expected stopping utility remains unchanged as all information is transmitted in equilibrium. 
The attention required by senders $i\in \{3, \cdots, n\}$ is also unaffected. The change in the receiver's payoff stems  from the attention required for component $\omega_{1,2}$.
Before the concentration, observing $\omega_1$ and $\omega_2$ requires expected attention cost of
$\E_{\boldsymbol{\w}_{-1}\sim \mu^0} [\bar v (\tilde \w_{1} \lvert \boldsymbol{\w}_{-1})] + \E_{\boldsymbol{\w}_{-2}\sim \mu^0} [\bar v (\tilde \w_{2} \lvert \boldsymbol{\w}_{-2})] $.
After the concentration, the new, merged, sender requires attention cost 
\begin{align*}
    \E_{\boldsymbol{\w}_{-\{1,2\}}\sim \mu^0} [\bar v (\tilde \w_{1,2} \lvert \boldsymbol{\w}_{-\{1,2\}})] = \E_{\boldsymbol{\w}_{-\{1,2\}}\sim \mu^0} \Brac{ \E_{\w_2 \sim \mu'(\boldsymbol{\w}_{-\{1,2\}})} [\bar v (\tilde \w_{1} \lvert (\boldsymbol{\w}_{-\{1,2\}},\w_2)) ] + \bar v (\tilde \w_{2} \lvert \boldsymbol{\w}_{-\{1,2\}}) },
\end{align*}
The first term on the right-hand side is equal to $\E_{\boldsymbol{\w}_{-1}\sim \mu^0} [\bar v (\tilde \w_{1} \lvert \boldsymbol{\w}_{-1})]$ by the law of iterated expectations. But $\E_{\boldsymbol{\w}_{-\{1,2\}}\sim \mu^0} \Brac{ \bar v (\tilde \w_{2} \lvert \boldsymbol{\w}_{-\{1,2\}}) }$ is greater than $\E_{\boldsymbol{\w}_{-2}\sim \mu^0} [\bar v (\tilde \w_{2} \lvert \boldsymbol{\w}_{-2})] $ by \eqref{eq:subst}.
Thus, concentrating the same amount of information on fewer senders slows down information transmission and hurts the receiver. 

For Corollary \ref{Cor:symmetry}, we apply Theorem \ref{AoN} to the Gaussian-quadratic specification in Example  \hyperref[ex:1]{1} to show that spreading a fixed amount of precision evenly across $n$ senders is optimal for the receiver. 
Define by $
     P \equiv p_0 +  \sum_{i=1}^n p_i,
$ the total precision, including the state's $p_0$ and each sender's $p_i$. 
In the AoN equilibrium sender $i$'s AoN probability is 
\begin{align*}
\lambda^*_i = c\frac{P(P-p_i)}{p_i}.
\end{align*}
The stopping utility is $-1/P$ when  $\boldsymbol{\w}_N$ is known and $-1/(P-p_i)$ when  $\boldsymbol{\w}_{-i}$ is known. 
A special feature of this setup is that the variance-reduction from each component is deterministic. As a result, the equilibrium rate  
$\lambda_i^*$ is constant. 
The receiver's final payoff is 
\begin{align*}
  - \frac{1}{P} - c \sum_{i=1}^n  \frac{1}{\lambda_i^*}  =  - \frac{1}{P}\left(1+  \sum_{i=1}^n  \frac{p_i}{ P-p_i} \right).
\end{align*}
\begin{Cor}[Symmetric senders] \label{Cor:symmetry}
Fixing the total and initial precision $P$ and $p_0$, the receiver's payoff is maximal if $p_i = (P-p_0)/n$ for all $i$. The symmetric allocation maximises the transmission rate.  
\end{Cor}

Fourth and finally, we amend the Gaussian-quadratic specification to consider the role of correlation among senders' components.
As before, let $\w_0 \sim \NN(0,1/p_0)$. There are conditionally i.i.d. signals $s_1, s_2, s_\text{c}$, with $s_k \sim \NN(\w_0, 1/p_k)$ for $k \in \{ 1,2,\text{c} \}$. 
There are $n=2$ senders with components 
$$\w_i = \frac{(1-\a)s_i + \a s_\text{c}}{\sqrt{(1-\a)^2 + \a^2}}, \quad \text{ for } i\in \{1,2\}. $$ The parameter $\a\in [0,1]$ determines the weight of the common signal $s_\text{c}$ in each component and thus the correlation between senders. The normalisation by $\sqrt{(1-\a)^2 + \a^2}$ ensures that the effect of the parameter $\a$ on informativeness of each component $\w_i$ and of $(\w_1,\w_2)$ jointly is determined by the relationship between $p_1$, $p_2$ and $p_\text{c}$ for all values of $\a \in [0,1]$. 
In particular, if the common component is sufficiently precise $p_\text{c} > p_1 + p_2$,  then the informativeness of $(\w_1,\w_2)$ about $\w_0$ is  increasing in $\a \in [0,1]$. 
Unsurprisingly, the receiver's equilibrium payoff also increases in $\a$ in this case. 
First, the receiver gets better information. Second, senders attract less attention because the marginal contribution of each $\w_i$ to the overall information decreases. 
The second effect creates a preference for correlated sources from the receiver's perspective even in cases in which correlation decreases the overall available information. 
\begin{Cor}[Benefit from correlated sources] \label{Cor:correlation}
Consider the AoN equilibrium of the Gaussian-quadratic setting above. There exists a threshold $  \bar p <  \min\{p_1, p_2\} $ 
such that 
the receiver's payoff under full correlation ($\a=1$) is strictly larger than under independence ($\a=0$) if  $p_\text{c} > \bar {p}$. 
\end{Cor} 
The calculations and exact value for $\bar p$ are in Appendix \ref{sec:app-proofs}.
Note that $p_c <  \min\{p_1, p_2\} $ implies that each component becomes less informative about $\w_0$. Corollary \ref{Cor:correlation} shows that even then the receiver may  benefit from correlation because the savings in attention costs exceed the informational loss.

\subsection{Large-market limit} \label{subsec:Gaussian}
When the number of senders $n$ grows large, the receiver learns the state exactly and at no attention cost. 
To leave the decision problem unchanged as $n$ changes, assume that $u$  depends on $\boldsymbol \omega$ only through the first entry $\omega_0$.
That is, $u(a,\boldsymbol{\w}) = u(a, \w_0)$ for all $a$ and all realisations $\boldsymbol{\w}_{N}$.
 Further, assume that all $\w_i$ for $i\in N$ are conditionally iid signals about $\w_0$. 
To avoid integer problems related to the exclusion of senders when Assumption \ref{Ass2} is violated, consider for this subsection the continuous-time analogue of the AoN equilibrium.\footnote{See footnote \ref{fn:exclusion} for alternative ways to rule out exclusion.} The belief $\mu^t$ is a right-continuous process taking value $\mu'(\boldsymbol{\w}_S) $ after  $\boldsymbol{\w}_S$ is revealed. If sender $i$'s information has not been revealed, she offers a Poisson process with revelation at rate $\l_i^*(\mu^t)$ and the receiver faces flow attention cost $c$ while consulting a sender. 

\paragraph{Competitive limit in Gaussian-quadratic example.}
First, we illustrate the competitive limit in the Gaussian-quadratic setting from Example 1. To make the signals identical, suppose that each source's precision is equal to $p_i=p>0$ for all $i\ge 1$.
As the number of senders $n$ goes to infinity, the receiver learns the state perfectly and for free:  
\begin{align}\label{eq:GaussQuadLimit}
\lim_{n\to \infty }  - \frac{1}{p_0 + n p } -    n c \frac{p}{c(p_0 + (n-1)p)(p_0+n p)} = 0.
\end{align}
The first term is the mean-squared error from the action after learning $n$ signals with precision $p$. The second term is the sum of attention costs the receiver spends at each sender. 
Thus, in the competitive limit no sender attracts any attention, and each sender's attention decreases fast enough so that the total attention cost of the receiver also vanishes. 
 
\paragraph{Competitive limit in general finite environments.}
We now show that the same conclusion, the receiver learns the state fully and for free in the limit, also holds for general finite environments. 
With Theorem \ref{AoN} at hand, this result can be shown rather directly using results in \cite{moscarini2002law}. We thus specify the finite environment parallel to their notation. 
Suppose there are $K<\infty$ actions $A =\{a_1, \dots, a_K\}$ and $M\le K$ payoff-relevant states $\Omega_0 ={\w_0^1, \dots, \w_0^M}$.  
Suppose further that for each realisation $\w^j_0$, action $a_j$ dominates all other actions. That is $u_k^j < u_j^j$ for all $k \ne j$ where $u_k^j = u(a_k, \w_0^j)$.
Each sender $i$'s component $\w_i$ is a conditionally independent draw from density $f(\cdot \vert \w_0) $. Assume that for any pair of realisations $\w^j_0 \ne \w^\ell_0$, the densities $f(\cdot \vert \w_0^j)$ and $f(\cdot \vert \w_0^\ell)$ are equivalent (mutually absolutely continuous) so that no signal realisation rules out any state. Assume further that $f(\cdot \vert \w_0^j) \ne f(\cdot \vert \w_0^\ell)$ so that each pair of states is distinguishable. 


The following result quantifies the precision of the receiver's belief after observing all $n$ signals and the sum of all senders' marginal contributions as $n$ grows large.
\begin{Pro}\label{prop:large-limit}
Consider the environment introduced above with conditionally iid signals $\w_i$ for $i\in \{1,\dots,n\}$. The state is learned perfectly in the limit:
\begin{align*}	
\lim_{n\to \infty} \E\Brac{ U(\mu'(\mu^0,\boldsymbol{\w})) \lvert  \w_0 } = U(\delta_{\w_0}) \;\; \text{a.s.}
\end{align*}
The sum of all senders' residual values vanishes:
\begin{align*}	
\lim_{n\to \infty} n \E_{\boldsymbol{\w}_{-i}}\Brac{ \bar v(\tilde \w_i \lvert \boldsymbol{\w}_{-i})} = 0.
\end{align*}
\end{Pro}

Again, the state is revealed perfectly and the required attention goes to 0 in the competitive limit. 
The proof applies the results in  \cite{moscarini2002law}  to show that the decision error and each sender's contribution decay exponentially as $n$ grows large. 
The limit in \eqref{eq:GaussQuadLimit} reveals that the finiteness of the environment is important for this approach in the general setting. In the Gaussian-quadratic setup with a continuum of states and actions, the attention each sender extracts decays only at quadratic rate as $n$ grows large.

\section{Concluding remarks} \label{sec:Concl}


This paper presents a parsimonious model to study dynamic information provision by senders who are interested in maximising attention.
A simple class of processes suffices to transmit all information from senders to the receiver with minimal attention. 
For the single sender case, the lack of intertemporal commitment paired with a finite action space makes jumps in the information process necessary. 
With competition, we identify a condition on the informational externalities that ensures that all information can be transmitted.


The model lends itself to several extensions that are beyond the scope of this paper. While we assume that all messages are publicly observable, modelling information as experiments can handle heterogeneous priors. Heterogeneous priors may arise if the senders do not observe the message of the visited competitor or if they do not observe the receiver's visit history. Incorporating such non-observabilities would allow a welfare comparison between the case in which information providers can track their users across sites and the case in which they are not permitted to do so. 

Other interesting avenues for future research are different aspects of information acquisition, such as the choice of issues to report on or the decision between seeking more or less correlation with other newspapers. The tractable computation of equilibrium payoffs in this model can be used as a reduced-form of the payoffs and applied to those questions. In an older working paper version \citep{knoepfle2020dynamic}, we consider an investigation race in which sources can choose how long to gather information before reporting and show that this exacerbates heterogeneity: sources who are more efficient in gathering information also investigate longer. 
Lastly, the introduction of prices in addition to attention allows for comparing membership-based business models to advertisement-based business models. 

The substitutes condition presents another avenue for future research. We present a condition directly expressed in terms of the valuation for information, which depends jointly on the decision problem and signal distributions. Characterising for a fixed decision problem, what signal distributions represent substitutable components remains an open question. Similarly, the effect of (partial) complements on information transmission presents a fruitful avenue. The coin-flip example illustrated how perfect complementary can thwart information transmission entirely in our setup. 
If components are partial complements such that each component is always worth some attention, but the value of each component increases with what has been learned from others, we conjecture that every equilibrium must extract the entire surplus from the receiver. 
Such equilibria under complementary information would necessarily exhibit some coordinated or `collusive' strategies by the senders where opponents respond to potential deviations with punishment strategies because, with complementarity, senders have an incentive to reveal their information later, when it is more valuable to the consumer.

Throughout the analysis we assume that senders have no preferences over the action taken by the receiver when he stops. The absence of a persuasion motive makes the analysis tractable. However, in many applications, for example when consultants or department heads provide advice, the senders likely have an interest in the final action. Our results go through if the senders share the receiver's action-state preferences but do not internalise the attention cost. Thus, the model can also be applied to department heads who want to be perceived as having influenced the CEO to take the 'right' action. 

\newpage

	\footnotesize	
	
	\setlength{\baselineskip}{1.9pt}
	
	\bibliographystyle{chicago}
	\bibliography{Bib.bib}

\begin{thebibliography}{}

\bibitem[\protect\citeauthoryear{Anderson and Renault}{Anderson and
  Renault}{2009}]{anderson2009comparative}
Anderson, S.~P. and R.~Renault (2009).
\newblock {Comparative Advertising: Disclosing Horizontal Match Information}.
\newblock {\em RAND Journal of Economics\/}~{\em 40\/}(3), 558--581.

\bibitem[\protect\citeauthoryear{Angeletos and Pavan}{Angeletos and
  Pavan}{2007}]{angeletos2007efficient}
Angeletos, G.-M. and A.~Pavan (2007).
\newblock {Efficient Use of Information and Social Value of Information}.
\newblock {\em Econometrica\/}~{\em 75\/}(4), 1103--1142.

\bibitem[\protect\citeauthoryear{Anton, Biglaiser, and Vettas}{Anton
  et~al.}{2014}]{anton2014dynamic}
Anton, J.~J., G.~Biglaiser, and N.~Vettas (2014).
\newblock {Dynamic Price Competition with Capacity Constraints and a Strategic
  Buyer}.
\newblock {\em International Economic Review\/}~{\em 55\/}(3), 943--958.

\bibitem[\protect\citeauthoryear{Au}{Au}{2015}]{au2015dynamic}
Au, P.~H. (2015).
\newblock {Dynamic Information Disclosure}.
\newblock {\em RAND Journal of Economics\/}~{\em 46\/}(4), 791--823.

\bibitem[\protect\citeauthoryear{Au and Whitmeyer}{Au and
  Whitmeyer}{2023}]{au2023attraction}
Au, P.~H. and M.~Whitmeyer (2023).
\newblock Attraction versus persuasion: Information provision in search
  markets.
\newblock {\em Journal of Political Economy\/}~{\em 131\/}(1), 202--245.

\bibitem[\protect\citeauthoryear{Ball}{Ball}{2023}]{ball2019dynamic}
Ball, I. (2023).
\newblock Dynamic information provision: Rewarding the past and guiding the
  future.
\newblock {\em Econometrica\/}~{\em 91\/}(4), 1363--1391.

\bibitem[\protect\citeauthoryear{Bergemann and Bonatti}{Bergemann and
  Bonatti}{2019}]{bergemann2019markets}
Bergemann, D. and A.~Bonatti (2019).
\newblock {Markets for Information: An Introduction}.
\newblock {\em Annual Review of Economics\/}~{\em 11}.

\bibitem[\protect\citeauthoryear{Bergemann, Bonatti, and Smolin}{Bergemann
  et~al.}{2018}]{bergemann2018design}
Bergemann, D., A.~Bonatti, and A.~Smolin (2018).
\newblock {The Design and Price of Information}.
\newblock {\em American Economic Review\/}~{\em 108\/}(1), 1--48.

\bibitem[\protect\citeauthoryear{Bergemann and Morris}{Bergemann and
  Morris}{2013}]{bergemann2013robust}
Bergemann, D. and S.~Morris (2013).
\newblock {Robust Predictions in Games With Incomplete Information}.
\newblock {\em Econometrica\/}~{\em 81\/}(4), 1251--1308.

\bibitem[\protect\citeauthoryear{Bergemann and V{\"a}lim{\"a}ki}{Bergemann and
  V{\"a}lim{\"a}ki}{2006}]{bergemann2006dynamic}
Bergemann, D. and J.~V{\"a}lim{\"a}ki (2006).
\newblock {Dynamic Price Competition}.
\newblock {\em Journal of Economic Theory\/}~{\em 127\/}(1), 232--263.

\bibitem[\protect\citeauthoryear{Billingsley}{Billingsley}{1995}]{billingsley1995probability}
Billingsley, P. (1995).
\newblock {\em {Probability and Measure}}.
\newblock Wiley Series in Probability and Statistics.
\newblock Wiley, 3rd ed.

\bibitem[\protect\citeauthoryear{Board and Lu}{Board and
  Lu}{2018}]{board2018competitive}
Board, S. and J.~Lu (2018).
\newblock {Competitive Information Disclosure in Search Markets}.
\newblock {\em Journal of Political Economy\/}~{\em 126\/}(5), 1965--2010.

\bibitem[\protect\citeauthoryear{B{\"o}rgers, Hernando-Veciana, and
  Kr{\"a}hmer}{B{\"o}rgers et~al.}{2013}]{borgers2013signals}
B{\"o}rgers, T., A.~Hernando-Veciana, and D.~Kr{\"a}hmer (2013).
\newblock {When are signals complements or substitutes?}
\newblock {\em Journal of Economic Theory\/}~{\em 148\/}(1), 165--195.

\bibitem[\protect\citeauthoryear{Che and Hörner}{Che and
  Hörner}{2017}]{che2017recommender}
Che, Y.-K. and J.~Hörner (2017, 12).
\newblock {Recommender Systems as Mechanisms for Social Learning}.
\newblock {\em Quarterly Journal of Economics\/}~{\em 133\/}(2), 871--925.

\bibitem[\protect\citeauthoryear{Che, Kim, and Mierendorff}{Che
  et~al.}{2023}]{che2019listener}
Che, Y.-K., K.~Kim, and K.~Mierendorff (2023).
\newblock {Keeping the Listener Engaged: A Dynamic Model of Bayesian
  Persuasion}.
\newblock {\em Journal of Political Economy\/}~{\em 131\/}(7), 1797--1844.

\bibitem[\protect\citeauthoryear{Che and Mierendorff}{Che and
  Mierendorff}{2019}]{che2019optimal}
Che, Y.-K. and K.~Mierendorff (2019).
\newblock {Optimal Dynamic Allocation of Attention}.
\newblock {\em American Economic Review\/}~{\em 109\/}(8), 2993--3029.

\bibitem[\protect\citeauthoryear{Chen, Eraslan, Ishida, and Yamashita}{Chen
  et~al.}{2024}]{chen2024optimal}
Chen, C.-H., H.~Eraslan, J.~Ishida, and T.~Yamashita (2024).
\newblock Optimal feedback dynamics against free-riding in collective
  experimentation.
\newblock {\em Working paper\/}.

\bibitem[\protect\citeauthoryear{Chen and Suen}{Chen and
  Suen}{2019}]{chen2019competition}
Chen, H. and W.~Suen (2019).
\newblock {Competition for Attention in the News Media Market}.
\newblock {\em Working paper\/}.

\bibitem[\protect\citeauthoryear{Choi, Dai, and Kim}{Choi
  et~al.}{2018}]{choi2018consumer}
Choi, M., A.~Y. Dai, and K.~Kim (2018).
\newblock {Consumer Search and Price Competition}.
\newblock {\em Econometrica\/}~{\em 86\/}(4), 1257--1281.

\bibitem[\protect\citeauthoryear{Dudey}{Dudey}{1992}]{dudey1992dynamic}
Dudey, M. (1992).
\newblock {Dynamic Edgeworth-Bertrand Competition}.
\newblock {\em The Quarterly Journal of Economics\/}~{\em 107\/}(4),
  1461--1477.

\bibitem[\protect\citeauthoryear{Ely and Szydlowski}{Ely and
  Szydlowski}{2020}]{ely2019moving}
Ely, J.~C. and M.~Szydlowski (2020).
\newblock {Moving the Goalposts}.
\newblock {\em Journal of Political Economy\/}~{\em 128\/}(2), 468--506.

\bibitem[\protect\citeauthoryear{Escud{\'e} and Sinander}{Escud{\'e} and
  Sinander}{2023}]{escude2023slow}
Escud{\'e}, M. and L.~Sinander (2023).
\newblock Slow persuasion.
\newblock {\em Theoretical Economics\/}~{\em 18\/}(1), 129--162.

\bibitem[\protect\citeauthoryear{Fudenberg, Strack, and Strzalecki}{Fudenberg
  et~al.}{2018}]{fudenberg2018speed}
Fudenberg, D., P.~Strack, and T.~Strzalecki (2018).
\newblock {Speed, Accuracy, and the Optimal Timing of Choices}.
\newblock {\em American Economic Review\/}~{\em 108\/}(12), 3651--84.

\bibitem[\protect\citeauthoryear{Fudenberg and Tirole}{Fudenberg and
  Tirole}{1991}]{fudenberg1991perfect}
Fudenberg, D. and J.~Tirole (1991).
\newblock {Perfect Bayesian Equilibrium and Sequential Equilibrium}.
\newblock {\em journal of Economic Theory\/}~{\em 53\/}(2), 236--260.

\bibitem[\protect\citeauthoryear{Fujishige and Yang}{Fujishige and
  Yang}{2003}]{fujishige2003note}
Fujishige, S. and Z.~Yang (2003).
\newblock A note on kelso and crawford's gross substitutes condition.
\newblock {\em Mathematics of Operations Research\/}~{\em 28\/}(3), 463--469.

\bibitem[\protect\citeauthoryear{Galperti and Trevino}{Galperti and
  Trevino}{2018}]{galperti2018shared}
Galperti, S. and I.~Trevino (2018).
\newblock {Shared Knowledge and Competition for Attention in Information
  Markets}.
\newblock {\em Working paper\/}.

\bibitem[\protect\citeauthoryear{Gossner, Steiner, and Stewart}{Gossner
  et~al.}{2021}]{gossner2021attention}
Gossner, O., J.~Steiner, and C.~Stewart (2021).
\newblock Attention please!
\newblock {\em Econometrica\/}~{\em 89\/}(4), 1717--1751.

\bibitem[\protect\citeauthoryear{Gul and Stacchetti}{Gul and
  Stacchetti}{1999}]{gul1999walrasian}
Gul, F. and E.~Stacchetti (1999).
\newblock Walrasian equilibrium with gross substitutes.
\newblock {\em Journal of Economic theory\/}~{\em 87\/}(1), 95--124.

\bibitem[\protect\citeauthoryear{Guo and Shmaya}{Guo and
  Shmaya}{2019}]{guo2019interval}
Guo, Y. and E.~Shmaya (2019).
\newblock {The Interval Structure of Optimal Disclosure}.
\newblock {\em Econometrica\/}~{\em 87\/}(2), 653--675.

\bibitem[\protect\citeauthoryear{H{\'e}bert and Zhong}{H{\'e}bert and
  Zhong}{2022}]{hebert2022engagement}
H{\'e}bert, B. and W.~Zhong (2022).
\newblock Engagement maximization.
\newblock {\em arXiv preprint arXiv:2207.00685\/}.

\bibitem[\protect\citeauthoryear{Jain and Whitmeyer}{Jain and
  Whitmeyer}{2020}]{jain2020competitive}
Jain, V. and M.~Whitmeyer (2020).
\newblock Competitive disclosure of information to a rationally inattentive
  consumer.
\newblock {\em Working paper\/}.

\bibitem[\protect\citeauthoryear{Kamenica and Gentzkow}{Kamenica and
  Gentzkow}{2011}]{kamenica2011bayesian}
Kamenica, E. and M.~Gentzkow (2011).
\newblock {Bayesian Persuasion}.
\newblock {\em American Economic Review\/}~{\em 101\/}(6), 2590--2615.

\bibitem[\protect\citeauthoryear{Ke and Lin}{Ke and
  Lin}{2020}]{ke2019informational}
Ke, T.~T. and S.~Lin (2020).
\newblock {Informational Complementarity}.
\newblock {\em Management Science\/}~{\em 66\/}(8), 3699--3716.

\bibitem[\protect\citeauthoryear{Kelso and Crawford}{Kelso and
  Crawford}{1982}]{kelso1982job}
Kelso, A. S.~J. and V.~P. Crawford (1982).
\newblock Job matching, coalition formation, and gross substitutes.
\newblock {\em Econometrica\/}, 1483--1504.

\bibitem[\protect\citeauthoryear{Knoepfle}{Knoepfle}{2020}]{knoepfle2020dynamic}
Knoepfle, J. (2020).
\newblock Dynamic competition for attention.
\newblock {\em Working paper\/}.
\newblock CRC TR224, University of Bonn and University of Mannheim.

\bibitem[\protect\citeauthoryear{Knoepfle and Salmi}{Knoepfle and
  Salmi}{2024}]{knoepfle2024dynamic}
Knoepfle, J. and J.~Salmi (2024).
\newblock Dynamic evidence disclosure: Delay the good to accelerate the bad.
\newblock {\em arXiv preprint arXiv:2406.11728\/}.

\bibitem[\protect\citeauthoryear{Koh and Sanguanmoo}{Koh and
  Sanguanmoo}{2022}]{koh2022attention}
Koh, A. and S.~Sanguanmoo (2022).
\newblock Attention capture.
\newblock {\em arXiv preprint arXiv:2209.05570\/}.

\bibitem[\protect\citeauthoryear{Koh, Sanguanmoo, and Zhong}{Koh
  et~al.}{2024}]{koh2024persuasion}
Koh, A., S.~Sanguanmoo, and W.~Zhong (2024).
\newblock Persuasion and optimal stopping.
\newblock {\em arXiv preprint arXiv:2406.12278\/}.

\bibitem[\protect\citeauthoryear{Kremer, Mansour, and Perry}{Kremer
  et~al.}{2014}]{kremer2014implementing}
Kremer, I., Y.~Mansour, and M.~Perry (2014).
\newblock Implementing the “wisdom of the crowd”.
\newblock {\em Journal of Political Economy\/}~{\em 122\/}(5), 988--1012.

\bibitem[\protect\citeauthoryear{Liang and Mu}{Liang and
  Mu}{2020}]{liang2019complementary}
Liang, A. and X.~Mu (2020).
\newblock {Complementary Information and Learning Traps}.
\newblock {\em The Quarterly Journal of Economics\/}~{\em 135\/}(1), 389--448.

\bibitem[\protect\citeauthoryear{Liang, Mu, and Syrgkanis}{Liang
  et~al.}{2022}]{liang2019dynamically}
Liang, A., X.~Mu, and V.~Syrgkanis (2022).
\newblock {Dynamically Aggregating Diverse Information}.
\newblock {\em Econometrica\/}~{\em 90\/}(1).

\bibitem[\protect\citeauthoryear{Margaria and Smolin}{Margaria and
  Smolin}{2018}]{margaria2018dynamic}
Margaria, C. and A.~Smolin (2018).
\newblock {Dynamic communication with biased senders}.
\newblock {\em Games and Economic Behavior\/}~{\em 110}, 330--339.

\bibitem[\protect\citeauthoryear{Mart{\'i}nez-de Alb{\'e}niz and
  Talluri}{Mart{\'i}nez-de Alb{\'e}niz and Talluri}{2011}]{martinez2011dynamic}
Mart{\'i}nez-de Alb{\'e}niz, V. and K.~Talluri (2011).
\newblock {Dynamic Price Competition with Fixed Capacities}.
\newblock {\em Management Science\/}~{\em 57\/}(6), 1078--1093.

\bibitem[\protect\citeauthoryear{Mayskaya}{Mayskaya}{2022}]{mayskaya2017dynamic}
Mayskaya, T. (2022).
\newblock {Dynamic Choice of Information Sources}.
\newblock {\em Working paper, California Institute of Technology\/}.

\bibitem[\protect\citeauthoryear{Morris and Shin}{Morris and
  Shin}{2002}]{morris2002social}
Morris, S. and H.~S. Shin (2002).
\newblock {Social Value of Public Information}.
\newblock {\em American Economic Review\/}~{\em 92\/}(5), 1521--1534.

\bibitem[\protect\citeauthoryear{Morris and Strack}{Morris and
  Strack}{2017}]{morris2017wald}
Morris, S. and P.~Strack (2017).
\newblock {The Wald Problem and the Equivalence of Sequential Sampling and
  Static Information Costs}.
\newblock {\em Working paper\/}.

\bibitem[\protect\citeauthoryear{Moscarini and Smith}{Moscarini and
  Smith}{2001}]{moscarini2001optimal}
Moscarini, G. and L.~Smith (2001).
\newblock {The Optimal Level of Experimentation}.
\newblock {\em Econometrica\/}~{\em 69\/}(6), 1629--1644.

\bibitem[\protect\citeauthoryear{Moscarini and Smith}{Moscarini and
  Smith}{2002}]{moscarini2002law}
Moscarini, G. and L.~Smith (2002).
\newblock The law of large demand for information.
\newblock {\em Econometrica\/}~{\em 70\/}(6), 2351--2366.

\bibitem[\protect\citeauthoryear{Orlov, Skrzypacz, and Zryumov}{Orlov
  et~al.}{2020}]{orlov2018persuading}
Orlov, D., A.~Skrzypacz, and P.~Zryumov (2020).
\newblock {Persuading the Principal to Wait}.
\newblock {\em Journal of Political Economy\/}~{\em 128\/}(7), 000--000.

\bibitem[\protect\citeauthoryear{Protter}{Protter}{2005}]{protter2005stochastic}
Protter, P.~E. (2005).
\newblock {\em Stochastic differential equations}.
\newblock Springer.

\bibitem[\protect\citeauthoryear{Radner and Stiglitz}{Radner and
  Stiglitz}{1984}]{radner1984nonconcavity}
Radner, R. and J.~Stiglitz (1984).
\newblock A nonconcavity in the value of information.
\newblock {\em Bayesian models in economic theory\/}~{\em 5}, 33--52.

\bibitem[\protect\citeauthoryear{Renault, Solan, and Vieille}{Renault
  et~al.}{2017}]{renault2017optimal}
Renault, J., E.~Solan, and N.~Vieille (2017).
\newblock {Optimal Dynamic Information Provision}.
\newblock {\em Games and Economic Behavior\/}~{\em 104}, 329--349.

\bibitem[\protect\citeauthoryear{Saeedi, Shen, and Shourideh}{Saeedi
  et~al.}{2024}]{saeedi2024getting}
Saeedi, M., Y.~Shen, and A.~Shourideh (2024).
\newblock Getting the agent to wait.
\newblock {\em arXiv preprint arXiv:2407.19127\/}.

\bibitem[\protect\citeauthoryear{Smolin}{Smolin}{2021}]{smolin2017dynamic}
Smolin, A. (2021).
\newblock Dynamic evaluation design.
\newblock {\em American Economic Journal: Microeconomics\/}~{\em 13\/}(4),
  300--331.

\bibitem[\protect\citeauthoryear{Sun}{Sun}{2011}]{sun2011disclosing}
Sun, M. (2011).
\newblock {Disclosing Multiple Product Attributes}.
\newblock {\em Journal of Economics \& Management Strategy\/}~{\em 20\/}(1),
  195--224.

\bibitem[\protect\citeauthoryear{Tauman, Urbano, and Watanabe}{Tauman
  et~al.}{1997}]{tauman1997model}
Tauman, Y., A.~Urbano, and J.~Watanabe (1997).
\newblock A model of multiproduct price competition.
\newblock {\em Journal of economic theory\/}~{\em 77\/}(2), 377--401.

\bibitem[\protect\citeauthoryear{Vives}{Vives}{1996}]{vives1996social}
Vives, X. (1996).
\newblock {Social Learning and Rational Expectations}.
\newblock {\em European Economic Review\/}~{\em 40\/}(3-5), 589--601.

\bibitem[\protect\citeauthoryear{Wald}{Wald}{1947}]{wald1947foundations}
Wald, A. (1947).
\newblock {Foundations of a General Theory of Sequential Decision Functions}.
\newblock {\em Econometrica\/}~{\em 15\/}(4), 279.

\bibitem[\protect\citeauthoryear{Wolinsky}{Wolinsky}{1986}]{wolinsky1986true}
Wolinsky, A. (1986).
\newblock {True Monopolistic Competition as a Result of Imperfect Information}.
\newblock {\em Quarterly Journal of Economics\/}~{\em 101\/}(3), 493--511.

\bibitem[\protect\citeauthoryear{Wu}{Wu}{2023}]{wu2023sequential}
Wu, W. (2023).
\newblock Sequential bayesian persuasion.
\newblock {\em Journal of Economic Theory\/}~{\em 214}, 105763.

\bibitem[\protect\citeauthoryear{Zhao, Mezzetti, Renou, and Tomala}{Zhao
  et~al.}{2024}]{zhao2024contracting}
Zhao, W., C.~Mezzetti, L.~Renou, and T.~Tomala (2024).
\newblock Contracting over persistent information.
\newblock {\em Theoretical Economics\/}~{\em 19\/}(2), 917--974.

\bibitem[\protect\citeauthoryear{Zhong}{Zhong}{2022}]{zhong2019optimal}
Zhong, W. (2022).
\newblock {Optimal Dynamic Information Acquisition}.
\newblock {\em Econometrica\/}~{\em 90\/}(4), 1537--1582.

\end{thebibliography}

	\newpage
	
	\numberwithin{equation}{section}

	\renewcommand\thesection{\Alph{section}}
\appendix
\section*{APPENDIX} 
	\setcounter{section}{0}
\setcounter{Le}{0}
     \renewcommand{\theLe}{\Alph{section}.\arabic{Le}}

\section{Updating of Information}	\label{AppProofs}


At the end of round $t$, If the receiver consults a sender in $t$, he uses message $m_t$ resulting from the selected experiment $\lambda_t$ to update the belief from $\mu^t$ to $\mu^{t+1}$. 
If, in the following expression, the denominator on the right-hand side is non-zero, the receiver forms $\mu^{t+1}$ through Bayes' rule as follows:
\begin{align*}
    \mu^{t+1}(\cdot) = \frac{ \underset{\cdot }{\int} \lambda_t(\omega_{d_t} , m_t)d\mu^t(\boldsymbol \w)}{ \underset{\Omega}{\int} \lambda_t(\omega_{d_t} , m_t)d\mu^t(\boldsymbol \w)}.
\end{align*}
In order to define the updating rule $\mu'(\mu^t, m)$ more generally, note that
$
    L^t(\cdot) \equiv \int_{\Omega } \lambda_t(\omega_{d_t}, \cdot) d\mu^t(\boldsymbol{\omega})
$
constitutes a probability measure over $M $. The updating rule $\mu'(\mu^t, m)$ is the non-negative function that satisfies
\begin{align*}
    &\int_{M'} \mu'(\mu^t,m)(\cdot) dL^{t}(m) = \int_{\cdot} \lambda_t(\omega_{d_t}, M')d\mu^t(\boldsymbol{\omega}),& &\text{ for all } M' \in \mathcal B (M).
\end{align*} Such a function $\mu'$ exists and is unique $L^t$-almost everywhere by the Radon-Nikodym Theorem, as for any $\Omega' \in \mathcal B(\Omega)$, the right-hand side, interpreted as a measure on $M$, is absolutely continuous with respect to $L^t$  \citep[see][ p. 422]{billingsley1995probability}.

 An experiment from sender $i$ contains information only about $\omega_i$ directly, i.e. $\lambda(\omega_i,\cdot)$ is independent of $\omega_j$ for all $j\ne i$. However, the receiver's belief about $\omega_j$ will still change through the correlation among components. If $d_t = i$, the likelihood ratio for two distinct realisations  $\boldsymbol{\omega}_{-i}$ and $\boldsymbol{\omega}_{-i}'$, given $\omega_i$ will not be changed through the updating. That is,
 $	\frac{\mu^{t+1}(\omega_i, \boldsymbol{\omega}_{-i}) }{\mu^{t+1}(\omega_i, \boldsymbol{\omega}'_{-i})} = \frac{\mu^t(\omega_i, \boldsymbol{\omega}_{-i}) }{\mu^t(\omega_i, \boldsymbol{\omega}'_{-i})}$  for all $\omega_i$.

Lastly, $v \ge 0 $ always, since 
\begin{align*}
U(\mu) &= \max_{a\in A} \E_{\boldsymbol\omega \sim \mu } \left[\; u(a,\boldsymbol\omega) \; \right] \\  & = \max_{a\in A}\E_{m \vert \mu} \left[ \E_{\boldsymbol\omega \sim \mu( \mu'(\mu, m))} \left[\; u(a,\boldsymbol\omega) \; \right]\right] \le  \E_{m\vert \mu} \left[ \max_{a\in A} \E_{\boldsymbol\omega \sim \mu( \mu'(\mu, m))} \left[\; u(a,\boldsymbol\omega) \; \right]\right],
\end{align*}
where I shorten the notation  from $\E_{\omega_i \sim \mu }  \E_{m \sim \lambda(\omega_i,\cdot)}$ to $\E_{m \vert \mu}$. The second equality is due to the martingale property of beliefs, which ensures that $\E_{m \vert \mu}\left[ \mu'(\mu, m)\right] = \mu$ for any experiment.

\section{Proofs} \label{sec:app-proofs}

\subsubsection*{Proof of Theorem \ref{thm:monopoly}}
Receiver optimality when faced with the sender's AoN strategy follows from the text after Theorem \ref{thm:monopoly}. 
Further, \begin{align*}
\frac{\bar v\left(\tilde \w_1 \right)}{c}    
\end{align*}
is clearly an upper bound for the expected rounds of attention. The AoN strategy ensures the sender this payoff, so she has no profitable deviation. 
Every equilibrium results in these payoffs because, increasing the AoN probability by $\epsilon>0$, the receiver strictly prefers accepting to rejecting, and the sender can ensure a payoff arbitrarily close to ${\bar v\left(\tilde \w_1 \right)}/ {c} $. Thus, there is no equilibrium that results in a strictly lower payoff than ${\bar v\left(\tilde \w_1 \right)}/ {c} $.
\qed

\subsubsection*{Proof of Theorem \ref{AoN}}

First, Assumption \ref{Ass2} and \ref{eq:subst} ensure that $\lambda_i^*(\mu^t)<1$ for all $i$ whose information has not been revealed at $t$.
The proof is organised in three claims:

\begin{Claim}
Take any strategies by senders $\ne\hspace*{-3pt} i$. Let the current belief be $\mu$ and assume sender $i$ has not revealed any information.
Suppose the receiver is playing a best response. 
Then, for any $\e> 0$ sender $i$ can assure an expected payoff of at least  $\frac{1}{\lambda_i^*(\mu)}(1-\e)$.
\end{Claim}
\textit{Proof of Claim 1.} 
The sender achieves this payoff by offering the increased AoN probability $\lambda_i^{(\e)}(\mu) = \frac{\lambda_i^*(\mu)}{1-\e}$, where we assume $\e < 1-  \lambda_i^*(\mu)$ so that $\l^{(\e)}_i(\m)<1$. Otherwise, the sender may lower $\e$ and achieve a higher payoff. 
First, we show that this AoN strategy ensures that the receiver will not stop without observing sender $i$'s information. Formally, consider an arbitrary history $h^t$ with current belief $\mu$, 
\begin{align*}
   \lambda^{(\e)}_i(\mu) E_{\omega_i\sim \mu} \left[V_R \left(h^t; (\l_{j,t})_{j\in N}, d_t=i, \w_i\right)\right] + (1- \lambda^{(\e)}_i(\mu))  V_R \left(h^t; (\l_{j,t})_{j\in N}, d_t=i, \emptyset\right) -c >  U\left(\mu\right).
\end{align*}
Here, $V_R$ denotes the receiver's continuation value if he chooses sender $i$ at time $t$. In the first term, sender $i$'s AoN experiment reveals $\w_i$. In the second term the AoN experiment reveals no information.  
Clearly, $V_R(h^{t+1})\ge U(\mu^{t+1})$, as the receiver always has the option to stop. 
To see that the 
above inequality is satisfied, consider the possible beliefs $\mu^{t+1}$, substitute $V^R$ on the left-hand side with $U$, and rearrange
\begin{align*}
 E_{\omega_i\sim \mu} \left[U \left(\mu'(\mu, \omega_i)\right)\right]  - U\left(\mu\right) \ge  \frac{c}{\lambda^{(\e)}_i(\mu) }. 
\end{align*}
By definition, the left-hand side of this inequality is exactly $\bar v\left(\tilde \w_i \ \vert \ \mu\right)$. Inserting for  $\lambda^{(\e)}_i$ on the right-hand side (using \eqref{eq:LamStar}) gives
\begin{align*}
    \bar v\left(\tilde \w_i \ \vert \  \mu \right) > (1-\e)\E_{\boldsymbol{\omega}_{-i} \sim \mu} \left[  \bar v\left(\tilde \w_i \ \vert \  \boldsymbol{\w}_{-i} \right)  \right],
\end{align*}
which holds because $\bar v$ satisfies \eqref{eq:subst}.

Given that the receiver eventually accepts $i$'s $\lambda_i^{(\e)}(\mu) $ offers, we show that the expected payoff for sender $i$ from using the AoN strategy is $\frac{1}{\lambda_i^{(\e)}(\mu)}$. 
The expected number of visits sender $i$ attracts is
\begin{align} \label{eq:visitsFormal}
  \E \left[  \sum_{n\ge 0} \prod_{m=1}^{n} \left( 1- \lambda^{(\e)}_i( \mu^{t(m)})  \right)  \right],
\end{align}
where $n$ counts the number of visits to sender $i$ and $t(m)$ is the round in which sender $i$ is visited the $m$'th time. The process $\left(\E \left[\left. \bar v (\tilde \w_i \ \vert \ \boldsymbol{\w}_{-i}) \; \right\vert \; \mathcal F_t \right]\right)_{t\ge 0}$ is a martingale. We write the sigma algebra $\mathcal F_t$ explicitly instead of the belief $\mu^t$ to cover the fact that different histories may lead to the same belief. By the strategy defined in \eqref{eq:LamStar} shows that $\lambda^{(\e)}$ is a convex function of this process. It follows that that the process $\left(\lambda^{(\e)}(\mu^t) \right)_{t\ge 0}$ is a submartingale. 
By receiver optimality, for any finite $m$, the stopping time $t(m)$ is finite almost surely. Further, $\lambda^{(\e)} \in [0,1]$, so that the submartingale has bounded increments. Thus, we can apply the optional sampling theorem \citep[see Theorem 35.2. in][]{billingsley1995probability} to derive that, for any $m' > m$: 
\begin{align*}
  1- \lambda^{(\e)}_i(\mu^{t(m)}) \ge   \E \left[ \left.\left( 1 - \lambda^{(\e)}_i(\mu^{t(m')}) \right) \right\vert \; \mathcal F_{t(m)} \right] .
\end{align*}
We now show how to derive the following lower bound for the number of visits in \eqref{eq:visitsFormal} given by 
\begin{align}\label{eq:submart}
    \sum_{n\ge 0} \left(1-\lambda_i^*(\mu^0)\right)^{n} = \frac{1}{\lambda_i^*(\mu^0)}.
\end{align}
To see how this is derived, consider, for illustration, the sum in \eqref{eq:visitsFormal} until $n = 2$, which satisfies 
\begin{align*}
&\E\left[ \left. 1 + \left(1- \lambda_i^*(\mu^{t(1) })\right)\left(1 + \left(1- \lambda_{i}^*(\mu^{t(2)}) \right) \right) \right\vert \; \mathcal{F}_{0}  \right] \\& \ge \E\left[ \left. 1 + \E \left[\left.  \left(1- \lambda_{i}^*(\mu^{t(2)}) \right) \right\vert \; \mathcal F_{t(1)} \right]\left(1 + \left(1- \lambda_{i}^*(\mu^{t(2)}) \right) \right) \right\vert \; \mathcal{F}_{0}  \right] \\
&= \E\left[ \left. 1 + \E \left[\left.  \left(1- \lambda_{i}^*(\mu^{t(2)}) \right) \right\vert \; \mathcal F_{t(1)} \right]\left(1 +\E \left[\left.  \left(1- \lambda_{i}^*(\mu^{t(2)}) \right) \right\vert \; \mathcal F_{t(1)} \right] \right) \right\vert \; \mathcal{F}_{0}  \right].
\end{align*}
The inequality uses \eqref{eq:submart} and the equality follows from the tower property of conditional expectations. 
This step can be reiterated. By Doob's martingale convergence theorem, the limit 
\begin{align*}
    \lim_{t\rightarrow \infty }   \E \left[ \left.\left( 1 - \lambda^{(\e)}_i(\mu^{t}) \right) \right\vert \; \mathcal F_{t(m)} \right]
\end{align*}
exists and is smaller or equal to $\left( 1 - \lambda^{(\e)}_i(\mu^{0}) \right)$. Applying these steps for all $n \in \N_{0}$, where the submartingale inequality and the tower property have to be used repeatedly for terms with $n>2$, gives the desired result.

\begin{Claim}
Let all senders $j \ne i$ play the AoN strategies from the theorem and assume the receiver is playing a best response. Then, it is optimal for sender $i$ to play the AoBN strategy from the theorem. 
\end{Claim}
\textit{Proof of Claim 2.}
By  Claim 1 the receiver visits all senders on path. Consider a history $h^t$ at which the components of senders $1,\cdots, j-1$ has been observed.\footnote{Any permutation of $j-1$ senders excluding $i$ works.} Then, the continuation utility of the receiver is 
\begin{align}\label{withi}
    V_R\left(h^t\right) = \E_{\boldsymbol{\omega} \sim \mu^t }\left[ U\left(\mu'(\boldsymbol{\omega})\right) \right] - c \sum_{k=j}^n V_k(\mu), 
\end{align}
where theprevious claim implies that the utility of sender $k$  from the AoN strategy is 
   $ V_k(\mu^t) = \frac{1}{\lambda_k^*(\mu^t)}$ for all  $k$.

Suppose now that one sender $i$ deviates to an offer that gives her an expected payoff higher than $\frac{1}{\lambda_{i}^*(\mu^t)}$ if accepted. 
By visiting the remaining, non-deviating senders, the receiver achieves an expected payoff of 
\begin{align} \label{withouti}
    \E_{\boldsymbol{\omega}_{-i} \sim \mu^t }\left[U\left(\mu'(\mu^0,\boldsymbol{\omega}_{-i})\right)\right] -  c \underset{ k\ne i}{\sum_{k=j }^n} \frac{1}{\lambda_k^*(\mu^t)}.
\end{align}
By construction of $\lambda_k^*$, at the last sender, the receiver is indifferent between stopping without her information or
paying the corresponding attention cost to obtain her information. Hence, the values \eqref{withi} and \eqref{withouti} are equal.
Thus,  the receiver  prefers to reject any offer yielding $i$ an expected payoff strictly higher than $\frac{1}{\lambda_{i}^*(\mu^t)}$, even if the deviating offer strategy of sender $i$ would lead to all her information being revealed.

\begin{Claim}
The AoN equilibrium achieves the maximal payoff for the receiver among all equilibria. 
\end{Claim}
\textit{Proof of Claim 3.}
The action is always taken with all information. As the value of information is always positive and, by claim 1, no sender can receive less attention in expectation, the AoN equilibrium is receiver-preferred. 
\qed

\subsubsection*{Calculations for Corollary \ref{Cor:correlation}}
Applying Theorem \ref{AoN}, the corollary follows from simple calculation of the expected mean-squared error in estimating $\w_0$ based on a single $\w_i$ or both $(\w_1,\w_2)$.

It is easily verified that 
\begin{align*}
    U(\mu^0) &= -\frac{1}{p_0}, 
    \\
U(\mu'(\w_i)) &= - \frac{\alpha ^2 p_i +(1-\alpha)^2 p_c }{p_0  \left(\alpha ^2 p_i +(1-\alpha )^2 p_c \right)+\left(\alpha ^2 +(1-\alpha )^2\right) p_i  p_c }     
  \\
U(\mu'(\w_1,\w_2)) &= -  \frac{\alpha ^2 (p_1 +p_2 )+(1-\alpha )^2 p_c }{ p_0\Paren{\alpha ^2  (p_1 +p_2 )+(1-\alpha )^2  p_c} +\left(\alpha ^2+(1-\alpha )^2\right)  (p_1 +p_2 ) p_c }. 
\end{align*}

The receiver's equilibrium payoff in the AoN equilibrium is $U(\mu'(\w_1,\w_2)) - \bar v(\tilde \w_1 \lvert \w_2 )/c - \bar v(\tilde \w_2 \lvert \w_1 )/c $, with $\bar v(\tilde \w_i \lvert \w_j )=  U(\mu'(\w_1,\w_2)) - U(\mu'(\w_j))$.  
Computing this payoff for $\a = 0$ and $\a=1$ reveals that 
$$ 
\bar  p = \frac{p_1 p_2 (2 p_0+p_1+p_2)}{(p_0+p_1)^2+p_2 (2 p_0+p_1+p_2)}.
$$

\subsubsection*{Proof of Proposition \ref{prop:large-limit}}
The fact that the receiver learns the state perfectly in the limit follows from the Strong Law of Large Numbers. Using the Chernoff bound one can show that the marginal contribution of each sender's information decreases exponentially as $n$ grows large, implying that the sum of all $n$ marginal contributions also vanishes. 
Under the additional assumptions of finite states and actions introduced before the result, we establish the two limits as direct consequences of results in \cite{moscarini2002law}. 

The convergence rates in Theorems 4 and 1 in \cite{moscarini2002law} imply, in the language of our setting, that
there exists a constant $0 < \kappa < \infty $ and a rate $\rho \in (0,1)$ such that  for $n$ large enough
\begin{align}\label{eq:expon-bound}
\E \Brac{U(\delta_{\{\w_0\}}) - U(\mu'(\m^0,\ (\w_1, \cdots, \w_n)))  } \le   \kappa  \rho^n. 
\end{align}

The first limit in Proposition \ref{prop:large-limit} follows immediately since $\E_{\w_0 \sim \mu }\Brac{U(\delta_{\{\w_0\}})} \ge U(\mu)$ for any belief $\mu$.

For the second inequality, note that $$ \bar v (\tilde \w_n \lvert (\w_1, \cdots, \w_{n-1})) =\E_{\w_n \sim \mu'(\mu^0,\ (\w_1, \cdots, \w_{n-1}))}\Brac{ U(\mu'(\mu^0,\ (\w_1, \cdots, \w_n)))} - U(\mu^0,\ (\w_1, \cdots, \w_{n-1}).$$
Since the stopping value after $n$ signals in the first term is always below the stopping value from knowing $\w_0$ precisely, we can bound the difference above by \eqref{eq:expon-bound} and conclude that 
that 
\begin{align*}
\E_{(\w_1, \cdots, \w_{n-1}) \sim \mu^0 }\Brac{ \bar v (\tilde \w_n \lvert (\w_1, \cdots, \w_{n-1})) } \le \kappa \rho^{n-1}.
\end{align*}
\qed

\section{Complements and Hold-Up Problem}\label{subsec:compl}

To illustrate how complementarities in the senders' information hinders information transmission to the sender, we present an example first discussed in Section \ref{subsec:AoN-eq} in which no information can be transmitted at all:
There are two senders $n=2$. 
Each sender observes the outcome of an independent, fair coin flip, $\W = \{H, T\}^2$ and $\mu^0(\w_1,\w_2) = 1/4$ for all $(\w_1,\w_2)\in \W$. The receiver has to guess whether the two coins match or not, $A = \{0, 1\}$. Let the receiver's utility again be 1 if he guesses correctly and 0 otherwise, $u(a,\boldsymbol{\w}) = a \mathbbm{1}\{\w_1=\w_2\} + (1-a)\mathbbm{1}\{\w_1\ne\w_2\}$. In this case, the two components (coin flips) are perfect complements. In particular, the value of observing one component without any information about the other is 0. 

Suppose that sender 1 has revealed the result of her coin flip. 
Sender 2 is a monopolist and requires $\frac{1}{c} \frac{1}{2} $ visits in expectation to reveal her information.\footnote{The value of sender 2's information after knowing $\omega_1$ is the difference between being able to guess correctly for sure or with probability $\frac{1}{2}$.} 
Anticipating this, the receiver's willingness to pay for sender 1's component is 0. The receiver is not willing to invest a single visit, even if sender 1 offers to reveal her information for sure. The cost $c>0$ is too high. 

There can be no information transmission in equilibrium due to this hold-up problem that arises with one sender after having observed the other sender's information. 

Going away from the case where the first sender offers to reveal her information perfectly, suppose that sender 1 revealed partial information, and for concreteness, let the current belief about $\omega_1$ be $\mu_1$ with $\frac{1}{2}<\mu_1<1$. Then, the value of sender 2's information is $\bar v( \tilde \w_2 \vert \mu_1) = \mu_1 - \frac{1}{2}$. After knowing the result of the second coin, the receiver guesses correctly whether they match or not with probability $\mu_1 >\frac{1}{2}$.  Sender 2 can extract at least $\bar  v( \tilde \w_2 \vert \mu_1) /c$ units of attention with the corresponding AoN strategy. Note that, as components are complements, the value of her information will increase in expectation with further revelations about $\omega_1$.  Thus, no information transmission is possible. This stylised result hinges on our modelling choices. If the receiver could wait for free, then information transmission could be supported, but the competing senders would extract all surplus.  Since the main focus of the analysis is on deriving conditions under which information transmission is possible, the model is stacked against transmission. To analysis how complementarities thwart information transmission, the results would be strengthened if some modelling choices would be changed to facilitate information transmission.

\end{document}